\documentclass[aps,nofootinbib,superscriptaddress, showpacs,preprintnumbers,  nofootinbibt,twocolumn]{revtex4-2}
\usepackage{epsfig}
\usepackage{multirow}
\usepackage{eurosym}
\usepackage{dcolumn}
\usepackage{bm}
\usepackage{enumerate}
\usepackage{float}
\usepackage{epstopdf}
\usepackage{amsmath}
\usepackage{bm}
\usepackage{amsfonts}
\usepackage{amssymb}
\usepackage{graphicx}
\usepackage{alphalph,mathtools}
\usepackage{etoolbox}
\usepackage{color}
\usepackage{booktabs}
\usepackage{footnote}
\usepackage{makecell,tabularx}
\usepackage{hyperref}
\hypersetup{colorlinks,citecolor=blue}
\hypersetup{colorlinks=true,linkcolor=red,filecolor=magenta,
    urlcolor=blue}

\def\be{\begin{equation}}
    \def\ee{\end{equation}}
\def\bea{\begin{eqnarray}}
    \def\eea{\end{eqnarray}}

\begin{document}

\title{\bf Observational Signatures of Modified Bardeen Black Hole: Shadow and Strong Gravitational Lensing }
\author{Niyaz Uddin Molla}
\email{niyazuddin182@gmail.com}\affiliation{Department of Mathematics, Indian Institute of Engineering Science and
Technology, Shibpur, Howrah-711 103, India,}
\author{Amna Ali}
\email{amnaalig@gmail.com} \affiliation{Department of Mathematics, Jadavpur University, Kolkata-700032, India,}
\author{Ujjal Debnath}
\email{ujjaldebnath@gmail.com}\affiliation{Department of Mathematics, Indian Institute of Engineering Science and
Technology, Shibpur, Howrah-711 103, India,}
\author{Saraswathy Shamini Gunasekaran}
\email{SShamini@uniten.edu.my} \affiliation{Institute Of Informatics $\&$ Computing In Energy (IICE),Universiti Tenaga Nasional (UNITEN), Malaysia.}

\begin{abstract}
    This study is dedicated to exploring the distinctive observational features affected by the modified Bardeen black hole through meticulous analysis of its shadow and strong gravitational lensing observations. We thoroughly investigated the impact of the black hole's parameters q, g, and $\mu$ on its shadow radius through numerical simulations and graphical representations. Notably, recent observations by the Event Horizon Telescope (EHT) scrutinized the images and shadows of supermassive black holes like $M87^*$ and $SgrA^*$. The observed angular diameters of the shadows were $\theta_d=42\pm3$ for $M87^*$ and $\theta_d=51.8\pm2.3$ for $SgrA^*$. Utilizing the EHT data, we constrained the parameters q and $\mu$ of the modified Bardeen black hole within specific ranges: $-0.89\leq \mu/8M^2 \leq 0.4$ and $0\leq |q|\leq 0.185$ for $M87^*$; and $-1.38\leq \mu/8M^2 \leq 0.1$ and $0\leq |q|\leq 0.058$ for $SgrA^*$, while maintaining the fixed value $g/2M=0.2$. This restriction of the ($\mu$, $q$) parameter space by the EHT findings illustrates the viability of modified Bardeen black holes as astrophysical candidates.

Additionally, we investigate gravitational lensing in the strong field limit for the modified Bardeen black hole, comparing its behavior to other astrophysical black holes such as the Schwarzschild ($\mu=\&q=0$) and regular Bardeen ($\mu=0$) black holes. Our investigation elucidated how the parameters of the modified Bardeen black hole influenced various observables associated with strong gravitational lensing. By examining the astrophysical ramifications through strong gravitational lensing, considering supermassive black holes in diverse galaxies, we uncovered that the modified Bardeen black hole exhibits distinct characteristics, offering a quantitative distinction from other black holes such as the Schwarzschild and regular Bardeen black holes. These findings in astrophysical consequences provide a promising pathway to differentiate the modified black hole from its counterparts in the realm of general relativity.

\end{abstract}



\maketitle
\section{Introduction} \label{sec:intro}


Black holes, postulated several decades ago within the framework of general relativity, stand as enigmatic and compact entities in our universe, pivotal to the foundation of astrophysics and high-energy physics \cite{zhang1991physics,de2012black,Nozari:2023flq}. These cosmic phenomena continue to captivate researchers across diverse scientific domains, owing to their profound implications on thermodynamics, quantum effects, and gravitational interactions within curved spacetime.

Recent years have witnessed a surge in both theoretical conjectures and substantial advancements in observational and experimental investigations surrounding black holes. Observational breakthroughs, notably the measurement of black hole spin in X-ray binaries, detection of gravitational waves from binary black hole mergers by LIGO \cite{LIGOScientific:2016aoc,LIGOScientific:2020iuh,LIGOScientific:2020stg}, and the unprecedented imaging of a black hole at the core of galaxy M87 by the Event Horizon Telescope (EHT) collaboration \cite{EventHorizonTelescope:2019dse}, have fortified our comprehension of these cosmic enigmas. Additionally, discoveries like the identification of wide star-black hole binary systems through radial velocity measurements \cite{Liu:2019lfc}have substantially contributed to reinforcing our understanding of black holes.

Beyond their significance in astronomy and astrophysics, black holes have spurred investigations in various branches of physics, particularly in addressing the singularity problem inherent in general relativity. The conception of regular black holes as solutions without singularities but with a horizon has paved the way for extensive studies on modified versions of classical black hole solutions. The Bardeen black hole, proposed as a regular alternative to the classical Schwarzschild black hole, addresses the singularity problem by replacing it with a de Sitter core \cite{bardeen1968proceedings}. Subsequent explorations led to variations like the Bardeen-anti-de Sitter and Bardeen-de Sitter black holes \cite{tzikas2019bardeen, fernando2017bardeen}, serving as crucial alternatives that extend our understanding beyond conventional paradigms.

The evolution of the Bardeen black hole has not been confined to its original formulation. Researchers have introduced modifications encompassing various physical effects and theoretical challenges \cite{Pourhassan:2015lfa,nag2023thermodynamics,Konoplya:2023ahd}, involving additional fields like scalar or electromagnetic fields and alterations to the theory of gravity itself, including higher-order curvature terms. Moreover, within 4D Einstein-Gauss-Bonnet (EGB) gravity, a modified Bardeen black hole presents fresh insights into regular black holes and their astrophysical implications \cite{islam2022strong}. Identifying a modified Bardeen black hole would profoundly impact our understanding of gravity, astrophysics, and fundamental physics, opening new avenues for exploration and enriching our knowledge of black holes' nature and properties.

Astronomical and astrophysical aspects, such as black hole parameter estimation, shadows, gravitational lensing, quasinormal modes, and particle motion around black holes, have been extensively studied for various types of black holes \cite{Virbhadra:1999nm, Rayimbaev:2022znx, Stuchlik:2019uvf,Boshkayev:2023rhr,Liu:2023kxd, Atamurotov:2023tff,Vishvakarma:2023csw,Rayimbaev:2022mrk, nag2023thermodynamics}.These investigations, carried out over the last few decades, have expanded our knowledge and understanding of black hole physics.

In this paper, we aim to build upon previous analyses conducted on Schwarzschild, modified regular, and regular Bardeen black holes (explored in the literature by \cite{Bozza:2002zj,Virbhadra:1999nm,Bozza:2003cp, Schee:2015nua, Eiroa:2010wm, Stuchlik:2019uvf, islam2022strong} and apply these insights to the modified Bardeen black hole. Our focus revolves around observing shadows and strong gravitational lensing, particularly investigating the astrophysical implications of the modified Bardeen black hole in comparison to other astrophysical black holes, such as the Schwarzschild and regular Bardeen black holes.

The black hole shadow, an intriguing astrophysical feature observed through strong gravitational lensing, has gained immense importance in recent years. The remarkable images of supermassive black holes in $M87^*$ and $SgrA^*$ released by the Event Horizon Telescope (EHT) collaboration \cite{EventHorizonTelescope:2019dse, EventHorizonTelescope:2019uob, EventHorizonTelescope:2019jan, EventHorizonTelescope:2019ths, EventHorizonTelescope:2019pgp, EventHorizonTelescope:2019ggy, EventHorizonTelescope:2022wkp, EventHorizonTelescope:2022apq, EventHorizonTelescope:2022exc, EventHorizonTelescope:2022urf, EventHorizonTelescope:2022xqj} provide pivotal evidence for the existence of black holes. These images reveal a distinct region at the heart of black holes, known as the black hole shadow. The concept of light rays bending due to gravitational lensing leads to the observation of a sharply defined boundary region of brightness on the distant image plane, offering robust tools for estimating black hole parameters \cite{Kumar:2018ple, ghosh2021parameters, Afrin:2021imp} and probing general relativity and its alternatives \cite{Mizuno:2018lxz, Psaltis:2018xkc, Stepanian:2021vvk, younsi2023black, Perlick:2021aok, walia2022testing, vagnozzi2022horizon,Hendi:2022qgi}.

Gravitational lensing, categorized into weak and strong regimes, serves as a powerful tool for understanding various properties of spacetime, the distribution of matter in the universe, and phenomena related to dark matter, dark energy, and cosmological parameters \cite{Hoekstra:2013via, Brouwer:2018xnj, Bellagamba:2018gec, Vanderveld:2012ec, Zhang:2021ygh, he2017direct, Cao:2012ja, Huterer:2017buf, Jung:2017flg, Andrade:2019wzn}. Our specific focus lies on strong gravitational lensing, which manifests as multiple images, arcs, and Einstein rings when a source is almost directly aligned behind a massive object.

The study of strong gravitational lensing holds immense potential for testing general relativity and alternative theories of gravity. Pioneering works by Bozza et al. \cite{Bozza:2001xd,Bozza:2002zj} laid the groundwork for obtaining the deflection angle in strong gravitational fields, shedding light on various types of black holes, naked singularities, and wormholes \cite{gao2021investigating,whisker2005strong, ghosh2021parameters,zhang2017strong,chen2009strong, eiroa2002reissner, petters2012singularity, virbhadra2002gravitational, sahu2012can, gyulchev2008gravitational, tsukamoto2021gravitational, paul2020strong, shaikh2019strong, tsukamoto2016strong, sharif2015strong}. These studies have extensively discussed astrophysical consequences, including the angular position, separation, magnification, Einstein rings, and time delays of relativistic images for rotating and non-rotating black holes \cite{kumar2022investigating, kumar2022testing, Kumar:2019pjp, kumar2020gravitational, afrin2023tests, Zhao:2017cwk, Ghosh:2023usx, Jha:2023qqz, Kuang:2022xjp, Chakraborty:2016lxo, Cavalcanti:2016mbe}.

Observationally, gravitational lensing phenomena by black holes have garnered significant attention, motivating our work to study the observational signatures of the modified Bardeen black hole through shadow and strong lensing observations. We intend to delve into various astrophysical consequences, including the black hole shadow, angular position, separation, Einstein ring, and time delays of relativistic images concerning the modified Bardeen black hole, and compare these with other astrophysical black holes, such as the Schwarzschild black hole and the ordinary regular Bardeen black hole.

The structure of this paper is organized as follows: In Section \ref{sec:modifiedBardeen}, we provide a concise overview of the modified Bardeen black hole, analyzing its null geodesics along the equatorial plane. Section \ref{sec:shadow} delves into studying the shadow of the modified Bardeen black hole and constraining its observables using observational data from $M87^*$ and $SgrA^*$. Following this, Section \ref{sec:Stronglensing} focuses on investigating strong gravitational lensing by the modified Bardeen black hole and exploring its observables, such as angular image position, separation, magnification, Einstein ring, and time delays of relativistic images. We also briefly overview the astrophysical implications of the modified Bardeen black hole. Subsequently, Section \ref{sec:comparison} estimates the strong lensing observables for the supermassive black hole BH $NGC 4649$ and compares these results with those obtained for the Schwarzschild black hole and the regular Bardeen black hole. Finally, in Section \ref{sec:results}, we discuss and summarize the outcomes of our study.

\section{ Modified Bardeen Black holes and Null geodesics}\label{sec:modifiedBardeen}

This paper explores the modified Bardeen black hole, initially conceptualized in \cite{Pourhassan:2015lfa}. This adaptation introduces alterations to the rotating iteration of the Bardeen black hole, exploring its potential as a mechanism for particle acceleration. Investigations into modified Bardeen black holes have been driven by diverse motivations, encompassing:
\begin{enumerate}
    \item Exploring the dynamics of particle acceleration involving two colliding particles in close proximity to the horizon of these black holes.
    \item Analyzing changes in the observable characteristics caused by the Bardeen black hole, particularly in relation to shadow effects and strong gravitational lensing.
    \item Investigating the thermodynamic attributes inherent in static Modified Bardeen black holes, specifically within the framework of Anti-de Sitter spacetime.
\end{enumerate}
These studies collectively contribute to a comprehensive understanding of modified Bardeen black holes, shedding light on their multifaceted implications across various scientific disciplines. This discussion focuses on the static version of the modified Bardeen black hole, as referenced in \cite{nag2023thermodynamics}, where we specifically examine the scenario with $a=0$. The static, spherically symmetric spacetime of this modified Bardeen black hole is mathematically described as presented in \cite{nag2023thermodynamics}.

\begin{equation}\label{1}
ds^2=-f(r)dt^2+ \frac{1}{h(r)} dr^2 +r^2 (d\theta^2 + \sin^2\theta d\phi^2)
\end{equation}
 where
\begin{equation}\label{1a}
  f(r) =\left(1-\frac{2Mr^2}{\left(q^2+r^2\right)^{3/2}}\right) \left(1-\frac{\mu M }{\left(g^2+r^2\right)^{3/2}}\right)
  \end{equation}

\begin{equation}\label{1b}
h(r)=  \left(1-\frac{2Mr^2}{\left(q^2+r^2\right)^{3/2}}\right)
\end{equation}

 The metric is defined by the magnetic charge $q$ and mass parameter $M$, while the parameters $\mu$ and $g$ serve to modify the spacetimes of the modified Bardeen black hole. This metric adheres to the following conditions:

i) It maintains Schwarzschild-like behavior as r becomes large.
ii) It incorporates the 1-loop quantum correction.
iii) It permits a finite time dilation between the central region and infinity.

When the parameter $\mu$ is absent, the metric (\ref{1}) simplifies to the regular Bardeen black hole. Additionally, setting $\mu=0$ and $q=0$ results in the Schwarzschild black hole.

The motion of photons around the modified Bardeen black hole is elucidated through the use of the Lagrangian formalism: $\mathcal{L}=-\frac{1}{2}g_{\mu \nu}\dot{x}^{\mu}\dot{x}^{\nu}$.

Without losing generality, we restricted the photon trajectory around the modified Bardeen black hole to the equatorial plane, where $\theta=\frac{\pi}{2}$. For the metric of the modified Bardeen spacetime (\ref{1}), the equation governing the motion of photons around the black hole is expressed as the Lagrangian equation:
  \begin{equation}\label{2}
  \begin{split}
 & \mathcal{L}=-\frac{1}{2}g_{\mu \nu}\dot{x}^{\mu}\dot{x}^{\nu}\\
 & =\frac{1}{2}(f(r)dt^2-\frac{1}{h(r)} dr^2 -r^2 (d\theta^2 + \sin^2\theta d\phi^2))=\delta\\
 \end{split}
  \end{equation}
  where $\dot{x}^{\mu}$ denotes the four-velocity of a photon, and the dot signifies differentiation with respect to the affine parameter $\tau$. The parameter $\delta$ takes on values of $-1$, $0$, or $1$, denoting spacelike, null, and timelike geodesics, respectively. Due to the metric coefficients in (\ref{2}), the coordinates $t$ and $\theta$ cannot be explicitly determined, leading to two conserved quantities: $E$ and $L$. These represent the energy and angular momentum of the particle, respectively, and can be defined as follows \cite{tsukamoto2017deflection,islam2022strong}: 

$E=-g_{\mu \nu}t^{\mu}\dot{x}^{\nu}=f(r)\dot{t}$ and $L=g_{\mu \nu}\phi^{\mu}\dot{x}^{\nu}=r^2\dot{\phi}$.

The photon travels along null geodesics around the modified Bardeen black hole, implying $\delta=0$. The null geodesics derived from equation (\ref{2}) are as follows:
   \begin{equation}\label{3}
   \dot{t}=\frac{dt}{d\tau}=\frac{E}{\left(1-\frac{2Mr^2}{\left(q^2+r^2\right)^{3/2}}\right) \left(1-\frac{\mu M }{\left(g^2+r^2\right)^{3/2}}\right)}
       \end{equation}

   \begin{equation}\label{4}
   \dot{\phi}= \frac{d\phi}{d\tau}=\frac{L}{r^2}
   \end{equation}

   \begin{equation}\label{5}
   \dot{r}=\frac{dr}{d\tau}= \pm\sqrt{h(r)\biggr(\frac{E^2 }{f(r)}-\frac{L^2}{r^2}\biggr)}
   \end{equation}
   The symbol $\pm$ in $\dot{r}$ signifies the direction of radial motion, distinguishing between ingoing and outgoing trajectories, respectively. The functions $f(r)$ and $h(r)$ are derived from Eqns. (\ref{1a}) and (\ref{1b}).

Equation (\ref{5}) can be expressed as
\begin{equation}\label{6}
   \biggr( \frac{dr}{d\tau}\biggr)^2+V_{eff}=0
\end{equation}
We can proceed without loss of generality by setting $E=1$. The effective potential function, denoted as $V_{\text{eff}}$, is described by:
\begin{equation}\label{7}
 V_{eff}=   h(r)\biggr(\frac{L^2}{r^2}-\frac{1 }{f(r)}\biggr)
\end{equation}
The critical photon ring orbit is characterized by the effective potential function, \(V_{\text{eff}}(r)\), satisfying specific conditions:
\[V_{\text{eff}}(r) = \frac{dV_{\text{eff}}(r)}{dr} = 0,\]
\[\frac{d^2V_{\text{eff}}(r)}{dr^2} > 0 \text{ (for stable)},\]
\[\frac{d^2V_{\text{eff}}}{dr^2} < 0 \text{ (for unstable)} \text{ circular orbit}.\]

\begin{figure*}[htbp]
\centering
\includegraphics[width=.45\textwidth]{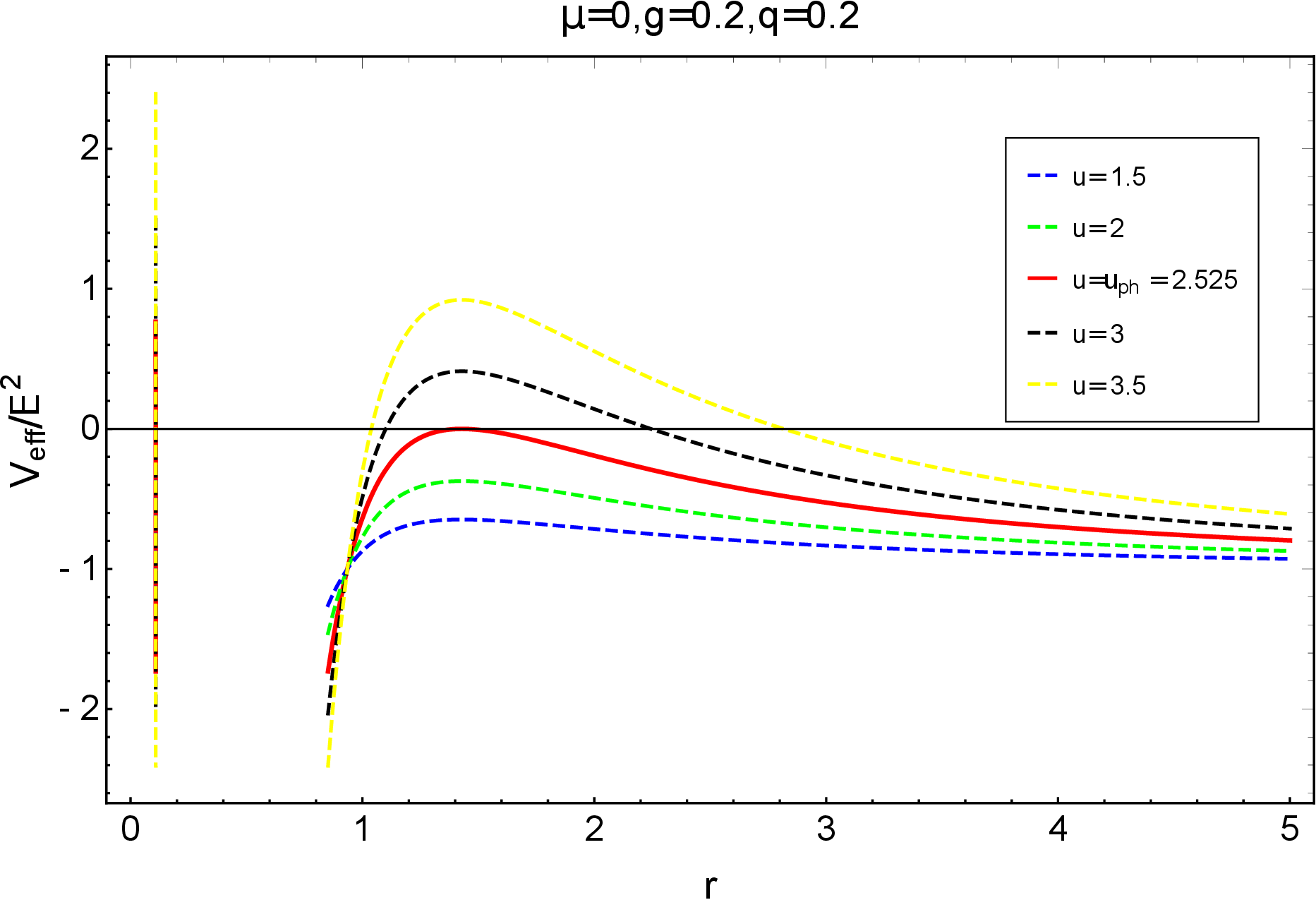}(a)
\qquad
\includegraphics[width=.45\textwidth]{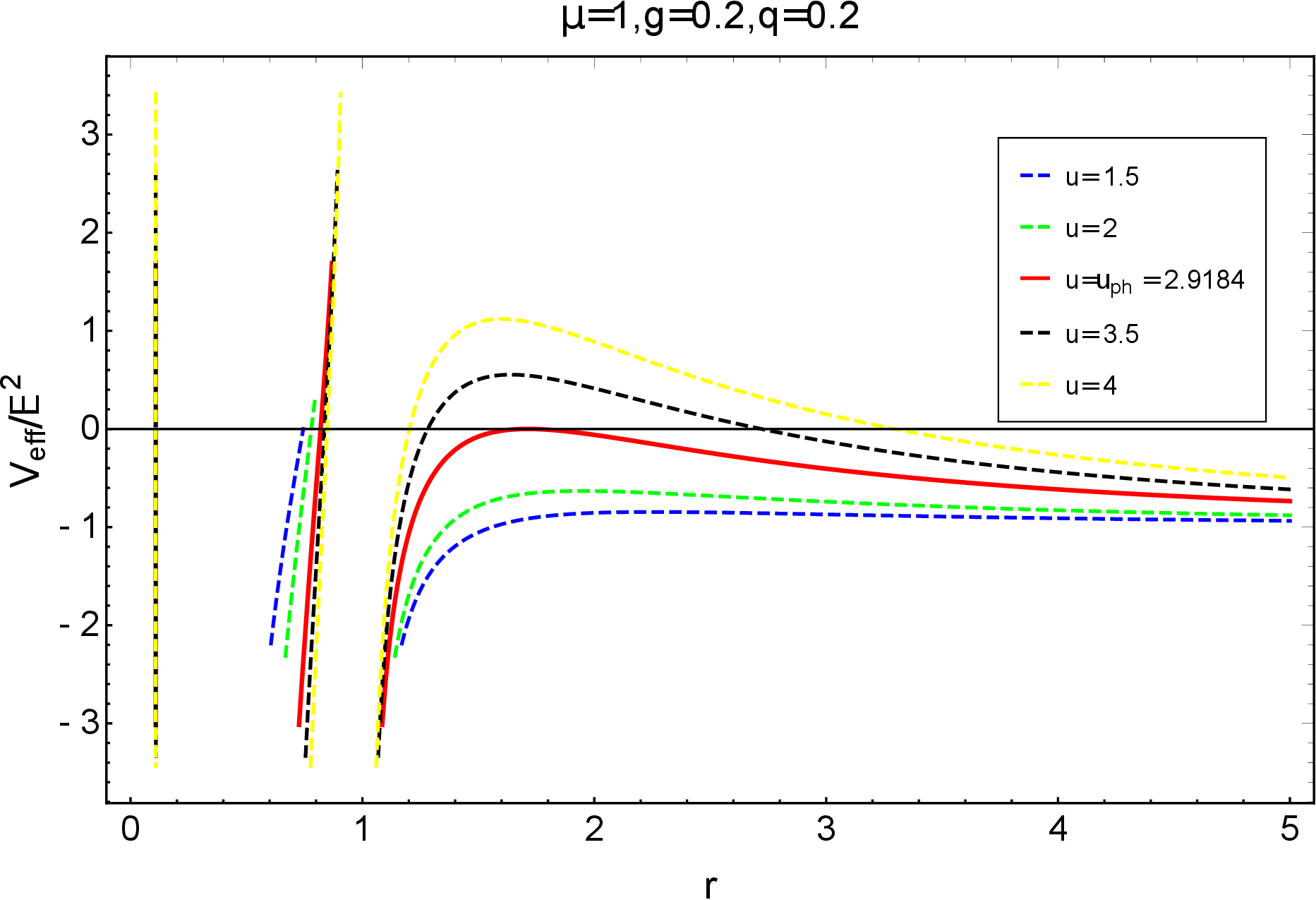}(b)
 \caption{Variation of the effective potential $V_{eff}$ for regular Bardeen (left panel) and modified Bardeen (right panel)black holes as a function of radial coordinate $r$.}
    \label{fig:1}
\end{figure*}

In the case of both the modified Bardeen and the ordinary regular Bardeen black hole, it's observed that \(\frac{d^2V_{\text{eff}}}{dr^2}|_{r_{\text{ph}}} < 0\), corresponding to an unstable circular photon orbit (refer to Fig.\ref{fig:1}). Consequently, photon rays arriving from infinity toward the vicinity of the modified Bardeen black hole, with a minimum impact parameter at the closest distance \(r_0\), follow unstable circular orbits around the black hole, forming a photon sphere with a radius of \(r_{\text{ph}}\).
\section{Shadows of modified Bardeen black hole}\label{sec:shadow}
The black hole's shadow stands as a crucial signature of the spacetime geometry, representing a dark area encompassed by circular orbits of photons forming what's known as a photon sphere. This phenomenon provides insight into the black hole's properties, optically revealing details related to gravitational lensing effects on nearby radiation. Several comprehensive reviews have delved into black hole shadows and their observables, offering extensive insights (see, for instance, \cite{Cunha:2018acu,Perlick:2021aok,Li:2020drn,Hu:2020usx,Hou:2022eev,Lee:2021sws,Saghafi:2022pme}). 

Notably, the Event Horizon Telescope (EHT) collaboration has utilized black hole shadow properties to detect images of black holes \cite{EventHorizonTelescope:2019dse,EventHorizonTelescope:2019pgp,EventHorizonTelescope:2019ggy}), garnering significant attention. In this section, we aim to explore the shadow of a modified Bardeen black hole and its observables, focusing on supermassive black holes like $M87^*$ and $SgrA^*$ positioned at the centers of galaxies.

The black hole's shadow is intimately linked to the critical impact parameter of the photon orbit. Utilizing the conditions discussed in the previous section, one can define the critical impact parameter:
\begin{equation}\label{8}
    u_{cr}=\frac{L}{E}=\frac{r_{ph}}{\sqrt{f(r_{ph})}}
    \end{equation}

where the  photon sphere radius  $r_{ph}$ is the largest real  root of the equation
     \begin{equation}\label{9}
    2f(r_{ph})-r_{ph} f^{\prime}(r_{ph})=0
     \end{equation}
The black hole shadow radius, denoted as \(r_{\text{sh}}\), while the observer is positioned far away from the black hole, can be expressed in terms of celestial coordinates (X, Y) as follows:
\begin{equation}\label{10}
    r_{sh}=\sqrt{X^2+Y^2}=\frac{r_{ph}}{\sqrt{f(r_{ph})}}
\end{equation}
where the celestial
co-ordinate ($X$, $Y$ ) at the boundary curve of black hole shadow defined as
\begin{equation}\label{11}
X=\lim_{r_0\rightarrow \infty}(r_0^2  \sin\theta_0)\frac{d\phi}{dr}
\end{equation}

\begin{equation}\label{12}
Y=\lim_{r_0\rightarrow \infty}(r_0^2  \frac{d\theta}{dr})
\end{equation}
Here, \(r_0\) represents the radial distance between the black hole and the observer, while \(\theta_0\) denotes the inclination angle between the observer and the black hole.

To express the shadow radius in terms of dimensional quantities, a set of transformations is applied: \(t \rightarrow \frac{t}{2M}\), \(r \rightarrow \frac{r}{2M}\), \(q \rightarrow \frac{q}{2M}\), \(g \rightarrow \frac{g}{2M}\), and \(\mu \rightarrow \frac{\mu}{8M^2}\) within the function \(f(r)\). This transformation defines the shadow radius as follows:
\begin{equation}\label{13}
    r_{sh}=r_{ph}\left[\left(1-\frac{r^2}{\left(q^2+r^2\right)^{3/2}}\right) \left(1-\frac{\mu }{\left(g^2+r^2\right)^{3/2}}\right)\right]^{-1/2}
\end{equation}
\begin{table*}
 \caption{Estimation of the photon sphere radius and the shadow radius for 
different values of black hole parameters .\label{table:1}}
\begin{tabular}{p{2.5cm}| p{1.5cm} |p{6cm}| p{6cm} }
\hline
\hline
\multicolumn{4}{c}{}\\
$\mu$ & $g$ & ~~~~Photon sphere radius $\mathbf{r_{ph}}$ & ~~~~~~Shadow radius $\mathbf{r_{sh}}$\\
\hline
 &  & $|q|=0.0$~~$|q|=0.1$~~$|q|=0.2$~~$|q|=0.4$ & $|q|=0.0$~~$|q|=0.1$~~$|q|=0.2$~~$|q|=0.4$ \\
 \tableline 
0 & 0 &$1.50000$~~$1.48309$~~$1.42899$~~$1.09915$ & $2.59808$~~$2.58054$~~$2.52504$~~$2.22043$ \\
\tableline 
1 & 0.2 &$1.74892$~~$1.73991$~~$1.71281$~~$1.60617$ & $2.95742$~~$2.94772$~~$2.91837$~~$2.79827$ \\
1& 1.2 &$1.58109$~~$1.56574$~~$1.51702$~~$1.24412$ & $2.79269$~~$2.77813$~~$2.73253$~~$2.50259$  \\
\tableline 
3 & 0.2 &$2.19323$~~$2.18977$~~$2.17956$~~$2.14160$ & $3.50634$~~$3.50177$~~$3.48813$~~$3.43500$  \\
3 & 1.2 &$1.84867$~~$1.83975$~~$1.81263$~~$1.69950$ & $3.21602$~~$3.20768$~~$3.18241$~~$3.07755$  \\
\tableline 
5 & 0.2 &$2.51788$~~$2.5158$~~$2.50965$~~$2.48650$ & $3.90482$~~$3.90178$~~$3.89272$~~$3.85741$  \\
5 & 1.2 &$2.15933$~~$2.15468$~~$2.14087$~~$2.08791$ & $3.60475$~~$3.59975$~~$3.58479$~~$3.52573$ \\
\tableline 
7 & 0.2 &$2.77523$~~$2.77374$~~$2.76930$~~$2.75243$ & $4.22334$~~$4.22334$~~$4.21412$~~$4.18715$  \\
7 & 1.2 &$2.43328$~~$2.43041$~~$2.42189$~~$2.38949$ & $3.93403$~~$3.93056$~~$3.92019$~~$3.87958$  \\
\hline 
\hline 
\end{tabular}
\end{table*}

Using Eqns. (\ref{9}) and (\ref{13}), we estimated both the photon sphere radius and the shadow radius across various values of black hole parameters: $\mu=0,1,3,5,7$; $g=0.2,1.2$; and $|q|=0,0.1,0.2,0.4$. These findings are summarized in Table \ref{table:1}. Notably, while keeping the parameters $\mu$ and $g$ constant, the photon sphere radius and the shadow radius of the black hole exhibit a decreasing trend as the parameter $q$ increases.

\subsection{Observational Constraints using  M87* and Sgr A* observations data}

We've previously investigated how the parameters $\mu$, $g$, and $q$ influence the photon sphere radius and the shadow radius. In this section, our goal is to determine the values of these parameters, $\mu$, $g$, and $q$, based on the observed angular diameter of the shadow. To simplify the parameter estimation process, we opt to fix the parameter $g=0.2$ and subsequently determine the potential range for the parameters $\mu$ and $q$.

For a distant observer, the measurement of a black hole's shadow image is consistently represented by the angular diameter $\theta_d$ \cite{Perlick:2021aok}. This angular diameter is given by:
\begin{equation}\label{14}
  \theta_d=  \frac{2u_{ph}}{D_{ol}}
\end{equation}
where $D_{ol}$ is the distance of the black hole to the observer.
The above equation can be expressed as
\begin{equation}\label{15}
   \theta_d  (\mu as)=  \biggr(\frac{6.191165 \times 10^6}{\pi}\biggr)(\frac{\gamma}{D_{ol}/Mps} )\biggr(\frac{2u_{ph}}{M}\biggr)
\end{equation}
where $\gamma$ denotes the mass ratio of a black hole to the sun, and $u_{ph}=u_{cr}$ is obtained from Eqn. (\ref{8}).

\begin{figure*}[htbp]
\centering
\includegraphics[width=.45\textwidth]{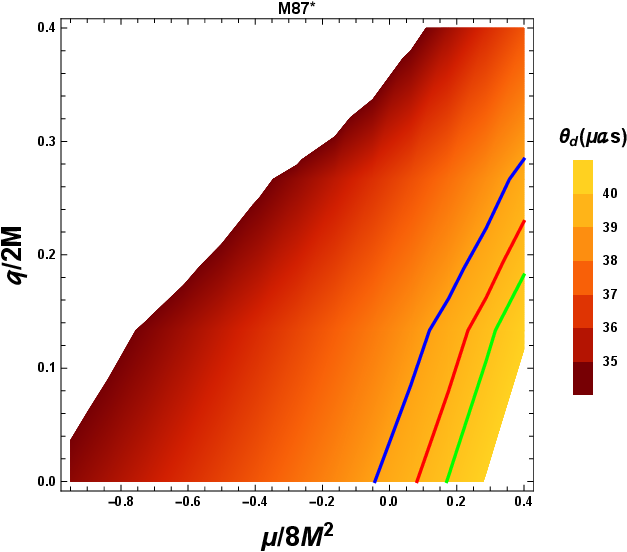}
\qquad
\includegraphics[width=.45\textwidth]{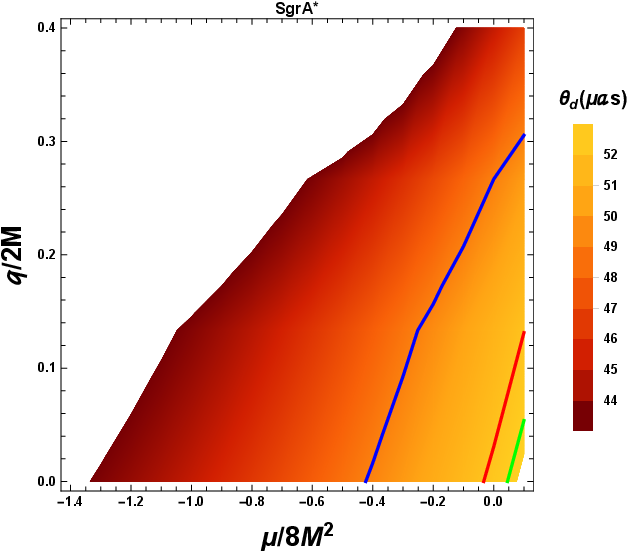}
  \caption{The angular diameter of the shadow for the modified Bardeen black hole is presented as a function of the parameters $\frac{\mu}{8M^2}$ and $\frac{q}{2M}$. In the left panel, the red solid curve corresponds to $\theta_d=39.4615 \mu as$ for the modified Bardeen black hole, the blue solid curve represents $\theta_d=38.80 \mu as$ for the Bardeen black hole, and the green solid curve corresponds to $\theta_d=39.9265 \mu as$ for the Schwarzschild black hole. These values fall within the $1\sigma$ region of the measured angular diameter, $\theta_d=42\pm3 \mu as$, for $M87^{*}$. In the right panel, for $SgrA^{*}$, the red solid curve represents $\theta_d=52.15 \mu as$ for the modified Bardeen black hole, the blue solid curve corresponds to $\theta_d=49.15 \mu as$ for the Bardeen black hole, and the green solid curve corresponds to $\theta_d=52.77 \mu as$ for the Schwarzschild black hole. These values also fall within the $1\sigma$ region of the measured angular diameter, $\theta_d=51.8\pm 2.3 \mu as$. }
    \label{fig:2}
\end{figure*}

By utilizing Eqns. (\ref{14}) and (\ref{15}), we've studied the angular diameter of the black hole shadow as a function of the parameters $\frac{\mu}{8M^2}$ and $\frac{q}{2M}$, presented in Fig.\ref{fig:2}. Our investigation into the angular diameter of the black hole shadow focuses on supermassive black holes, specifically $M87^*$ with mass and distance from Earth approximately $M\approx 6.5 \times 10^9 \dot{O}$ and $D_{ol}\approx 16.8 \text{ Mpc}$ respectively, and $Sgr A^*$ with mass and distance from Earth approximately $M\approx 4.28\times 10^6\dot{O}$ and $D_{ol}\approx 8.32\text{ kpc}$ respectively. The measured shadow angular diameters are $\theta_d=42\pm3$ for $M87^*$ and $\theta_d=51.8\pm2.3$ for $Sgr A^*$ \cite{EventHorizonTelescope:2022wkp,EventHorizonTelescope:2022apq}.

We've constrained the modified black hole parameters $q$ and $\mu$ while keeping the value of $g$ fixed based on EHT collaboration data for the angular shadow diameter of $M87^*$ and $SgrA^*$. Our observations indicate constrained ranges for the parameters $\mu$ and $q$ of the modified Bardeen black hole: $-0.89\leq \frac{\mu}{8M^2} \leq 0.4$ and $0\leq |q|\leq 0.185$ for $M87^*$; and $-1.38\leq \frac{\mu}{8M^2} \leq 0.1$ and $0\leq |q|\leq 0.058$ for $SgrA^*$, while maintaining the fixed value of $\frac{g}{2M}=0.2$. These additional parameters $\mu$, $g$, and $q$ characterize the modified Bardeen black holes akin to the supermassive black holes $M87^*$ and $SgrA^*$. Notably, our results suggest that the modified Bardeen black hole complies with the EHT constraint, signifying its potential detectability and differentiation from other astrophysical black holes in future observations.


\section{Strong gravitational lensing  and it's observable}\label{sec:Stronglensing}
We aim to explore the effects of strong gravitational lensing caused by the modified Bardeen black hole, investigating how its parameters—$\mu$, $q$, and $g$—influence several astrophysical observations. Our focus includes analyzing effects such as changes in angular position, separation, magnification, the formation of Einstein's rings, and time delays of relativistic images. We'll compare these observations with both an ordinary regular Bardeen black hole ($\mu=0$) and a standard Schwarzschild black hole ($\mu=0$, $q=0$).

In our investigation, we're particularly interested in determining the strong deflection angle of photon rays within the equatorial plane ($\theta=\frac{\pi}{2}$) due to the modified Bardeen black hole. To compute this angle, we apply dimensionless operations on metric (\ref{1}): $t\rightarrow\frac{t}{2M}$, $r\rightarrow\frac{r}{2M}$, $q\rightarrow\frac{q}{2M}$, $g\rightarrow\frac{g}{2M}$, and $\mu\rightarrow\frac{\mu}{8M^2}$. This allows us to better understand the impact of these parameters on the gravitational lensing effects near the black hole. Rewriting the metric:
 \begin{equation}\label{16}
d\bar{s}^2=-A(x)dt^2+ B(x) dr^2 +C(x) d\phi^2
\end{equation}
where
$$
  A(r) =\left(1-\frac{r^2}{\left(q^2+r^2\right)^{3/2}}\right) \left(1-\frac{\mu }{\left(g^2+r^2\right)^{3/2}}\right)$$

$$B(r)=  \frac{1}{\left(1-\frac{r^2}{\left(q^2+r^2\right)^{3/2}}\right)}$$

  and $$  C(r)=r^2 $$

 When particles approach the closest distance $r=r_0$ to the central black hole, signifying $\frac{dr}{d\tau}=0$, we can define the minimum impact parameter $u_0$ in relation to the closest distance $r_0$ \cite{Bozza:2002zj} as:
  \begin{equation}\label{17}
   u_0=\frac{r_0}{\sqrt{A(r_0)}}
   \end{equation}

\begin{figure*}[htbp]
\centering
\includegraphics[width=.45\textwidth]{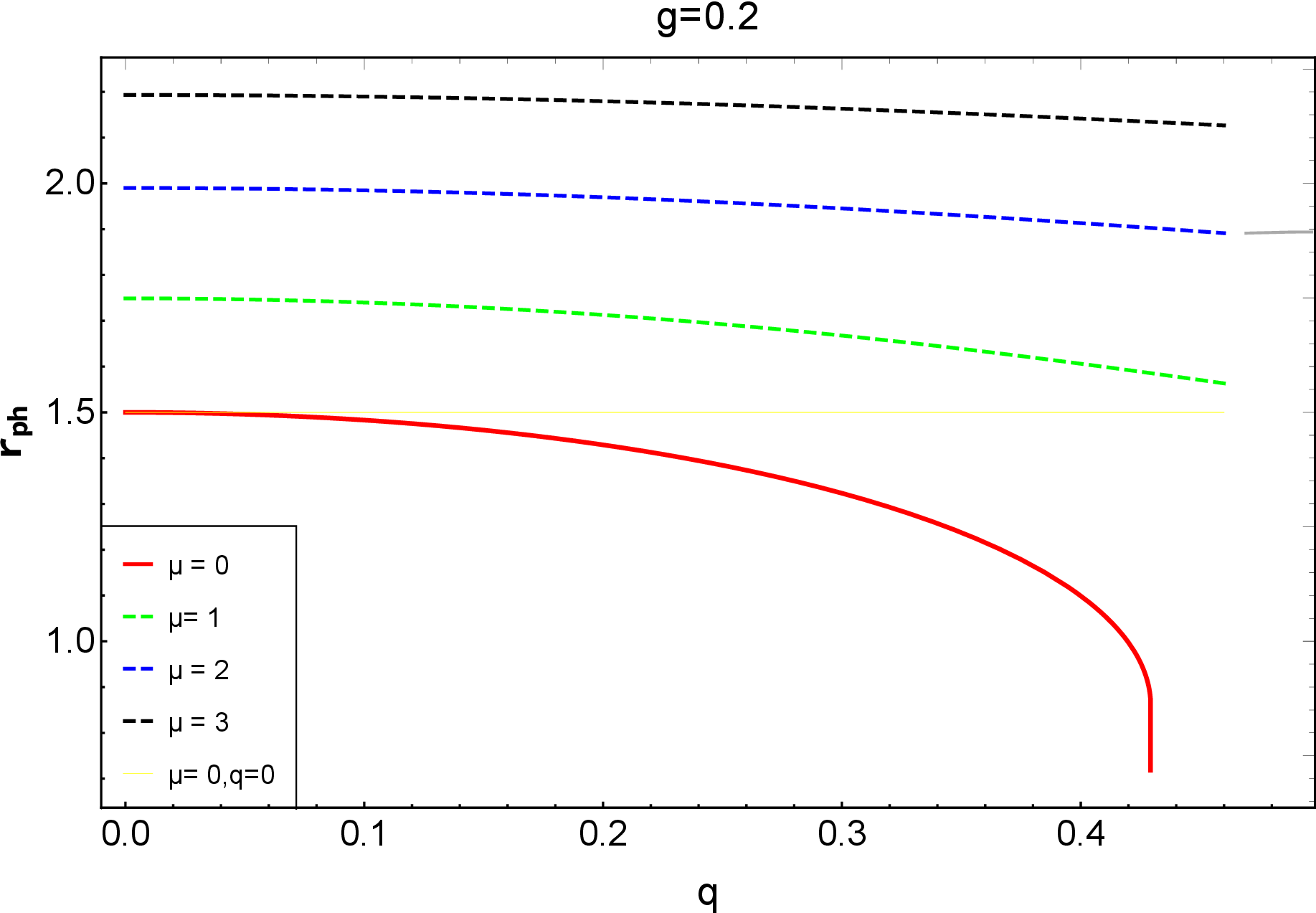}(a)
\qquad
\includegraphics[width=.45\textwidth]{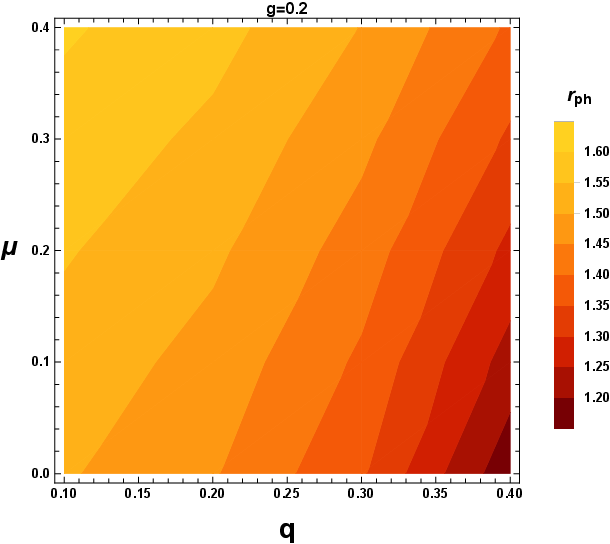}
 \caption{The behaviour of the   photon sphere radius $\mathit{r_{ph}}$ vs  parameter $q$ with the different values of $\mu$ for fixed value of $g=0.2$ (left panel) ;and the photon sphere radius $\mathit{r_{ph}}$ as a function of both the parameters $\mu$ and $q$ for the fixed value of $g=0.2$ (right panel) .}
    \label{fig:3}
\end{figure*}

\begin{figure*}[htbp]
\centering
\includegraphics[width=.45\textwidth]{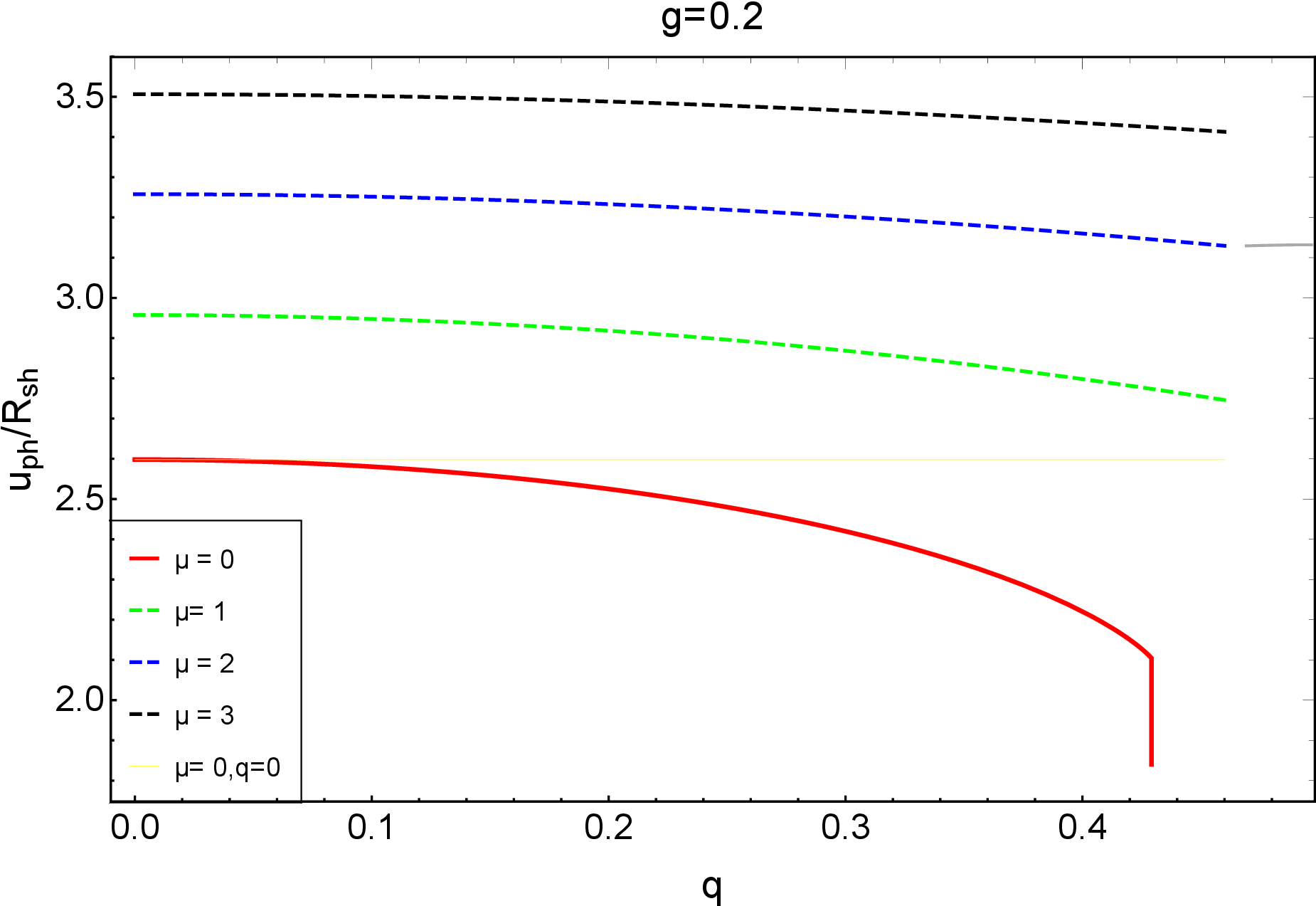}(a)
\qquad
\includegraphics[width=.45\textwidth]{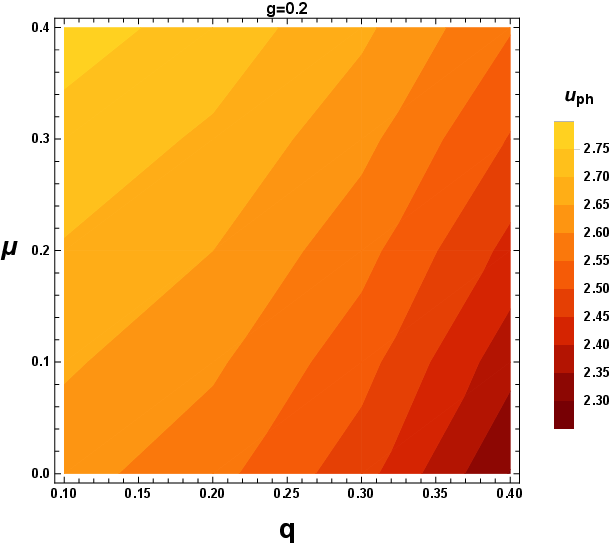}(b)
 \caption{The behaviour of the   minimum impact parameter $\mathit{u_{ph}/R_{sh}}$ vs  parameter $q$ with the different values of $\mu$ for fixed value of $g=0.2$ (left panel) ;and the   minimum impact parameter $\mathit{u_{ph}}/R_{sh}$ as a function of both the parameters  $\mu$ and $q$ for the fixed value of $g=0.2$ (right panel) .}
    \label{fig:4}
\end{figure*}

The behavior of the photon sphere radius, \(r_{ph}\), has been depicted as a function of the parameter \(q\) in Fig.\ref{fig:3}(a), and as a function of the parameters \(\mu\) and \(q\) in Fig.\ref{fig:3}(b). Notably, it has been observed that, while keeping the parameters \(\mu\) and \(g\) fixed, the photon radius \(r_{ph}\) slightly decreases with the parameter \(q\). Conversely, for fixed parameters \(g\) and \(q\), the photon radius \(r_{ph}\) tends to increase with the parameter \(\mu\).

Moreover, in Fig.\ref{fig:3}(a), the photon radius \(r_{ph}\) corresponding to the modified Bardeen black hole is observed to exceed that of the ordinary regular Bardeen black hole (red solid line). Additionally, this \(r_{ph}\) value for the modified Bardeen black hole surpasses the value of \(r_{ph}=1.5\), which corresponds to the Schwarzschild black hole (yellow horizontal line) \cite{Bozza:2002zj}.

When $r_0 \rightarrow r_{ph}$,  deflection angle becomes divergent  and for
 $r_0 >r_{ph}$, it becomes finite only. The photon having impact parameter $u<u_{ph}$ falls into the black hole and for the case when $u>u_{ph}$, it reaches the closest distance $r_0$ near the black hole; while for the impact parameter $u=u_{cr}=u_{ph}$, photon revolved in unstable orbit around the black hole.

The critical impact parameter for the unstable photon orbit, \(u_{ph}\), is determined by the equation:
\begin{equation}\label{18}
u_{ph} =\frac{r_{ph}}{\sqrt{A(r_{ph})}}
 \end{equation}
Its behavior has been graphically represented as a function of the parameter \(q\) in Fig.\ref{fig:4}(a), and as a function of the parameters \(\mu\) and \(q\) in Fig.\ref{fig:4}(b). Upon examination, it has been noted that for fixed parameters \(\mu\) and \(g\), the critical impact parameter \(u_{ph}\) experiences a slight decrease with the parameter \(q\). Conversely, for fixed parameters \(g\) and \(q\), \(u_{ph}\) tends to increase with the parameter \(\mu\) (refer also to Table.3).

In Fig.\ref{fig:4}(a), it's further observed that the critical impact parameter \(u_{ph}/R_s\) corresponding to the modified Bardeen black hole surpasses that of the ordinary regular Bardeen black hole (red solid line). Additionally, this \(u_{ph}/R_s\) value for the modified Bardeen black hole exceeds the value of \(u_{ph}/R_{sh}=2.59808\), which corresponds to the Schwarzschild black hole (yellow horizontal line) \cite{Bozza:2002zj}.

The  strong deflection angle
for the modified Bardeen black hole spacetime, as a function of  the closest approach distance  $r_0$, can be read as: \cite{Virbhadra:2002ju,Claudel:2000yi,Zhao:2017cwk})

\begin{equation}\label{19}
\alpha_D(r_0)=I(r_0)-\pi=2\int_{r_0}^\infty \frac{\sqrt{B(r)}dr}{\sqrt{C(r) \sqrt{ \frac{A(r_0)C(r)}{A(r)C(r_0)}-1} }} dr-\pi
\end{equation}

 The strong deflection angle $\alpha_D(r_0)$  depends upon the relation between $r_0$ and $r_{ph}$ and while $r_0\approx r_{ph}$, it is increased.
 So, we define a new variable z as:
 \begin{equation}\label{20}
 z=1-\frac{r_0}{r}
 \end{equation}

For,  $r_0\approx r_{ph}$, the strong  deflection angle becomes:
  \begin{equation}\label{21}
     \alpha_D(u)= -\bar{a}~ log\left(\frac{u}{u_{ph}}-1\right) +\bar{b} +\mathcal{O}((u -u_{ph})log(u -u_{ph}))
     \end{equation}
    where

   \begin{equation}\label{22}
       \bar{a}
       =\sqrt{\frac{2A(r_{ph}B(r_{ph})}{A(r_{ph})C^{\prime\prime}(r_{ph})-A^{\prime\prime}(r_{ph}C(r_{ph})}}
   \end{equation}
   and
   \begin{equation}\label{23}
      \bar{b}=-\pi + I_R(r_{ph}+ \bar{a}~ log \biggr[r_{ph}^2\biggr(\frac{C^{\prime\prime}_{ph}}{c_{ph}}-\frac{A^{\prime\prime}_{ph}}{A_{ph}}\biggr)\biggr],
  \end{equation}

Here,
  $ I_R(r_{ph})= 2 \int_{0}^{1} \Bigg(r_{ph}\biggr[\sqrt{\frac{B(z)}{C(Z)}}\biggr(\frac{A(r_{ph})}{C(r_{ph})}\frac{C(z)}{A(z)}-1\biggr)\frac{1}{(1-z)^2} \biggr]-\frac{\bar{a}}{z ~r_{ph}}\Bigg)dz  $, which
  is obtained numerically.

\begin{figure*}[htbp]
\centering
\includegraphics[width=.45\textwidth]{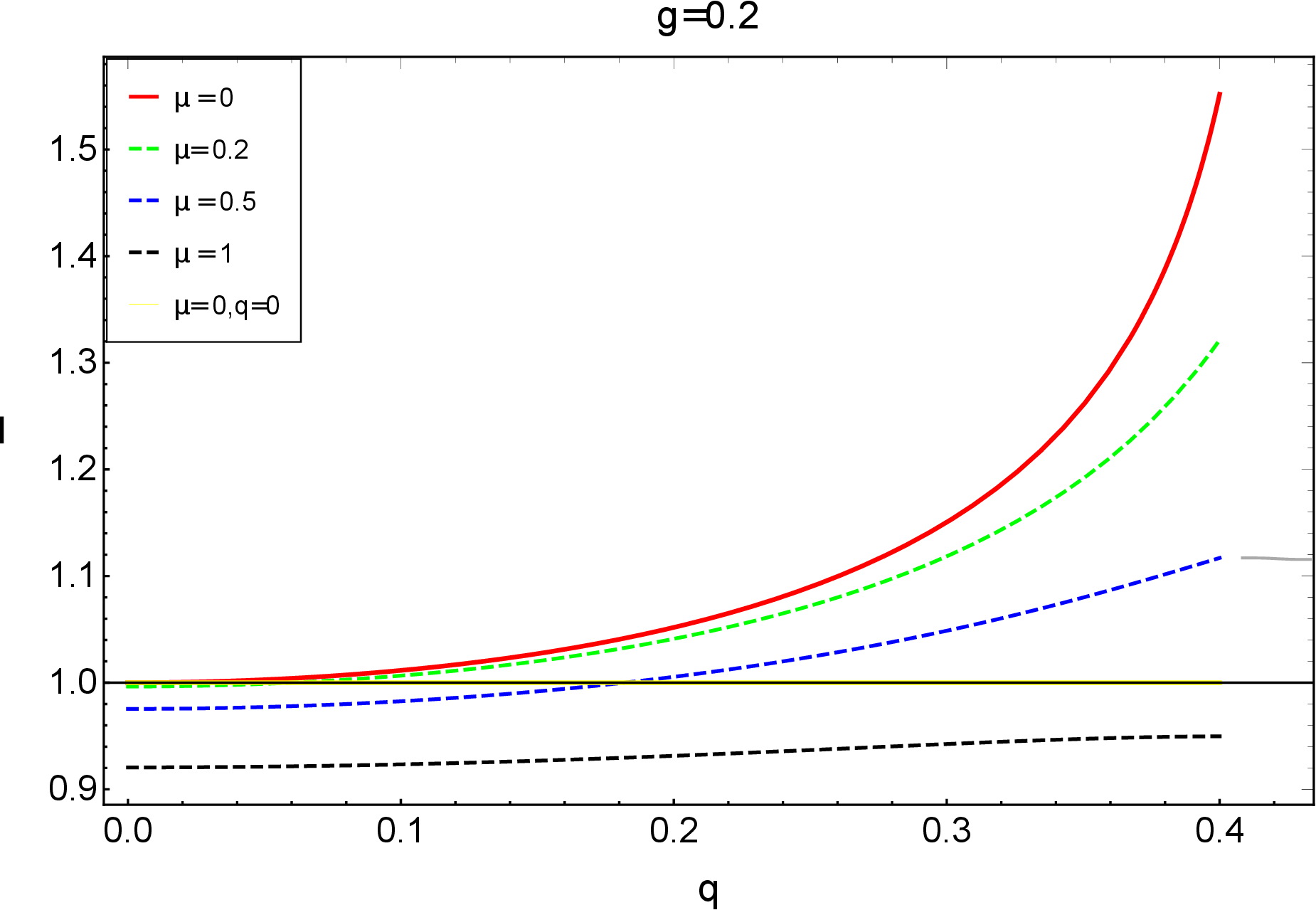}
\qquad
\includegraphics[width=.45\textwidth]{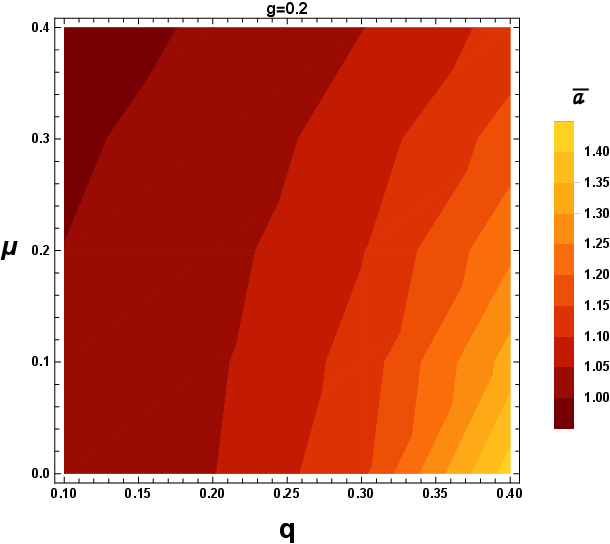}
  \caption{The behaviour of the   deflection limit coefficient  $\mathit{\bar{a}}$ vs  parameter $q$ with the different values of $\mu$ for fixed value of $g=0.2$ (left panel) ;and the   deflection limit coefficient  $\mathit{\bar{a}}$ as a function of both the parameters  $\mu$ and $q$ for the fixed value of $g=0.2$ (right panel) .}
    \label{fig:5}
\end{figure*}

\begin{figure*}[htbp]
\centering
\includegraphics[width=.45\textwidth]{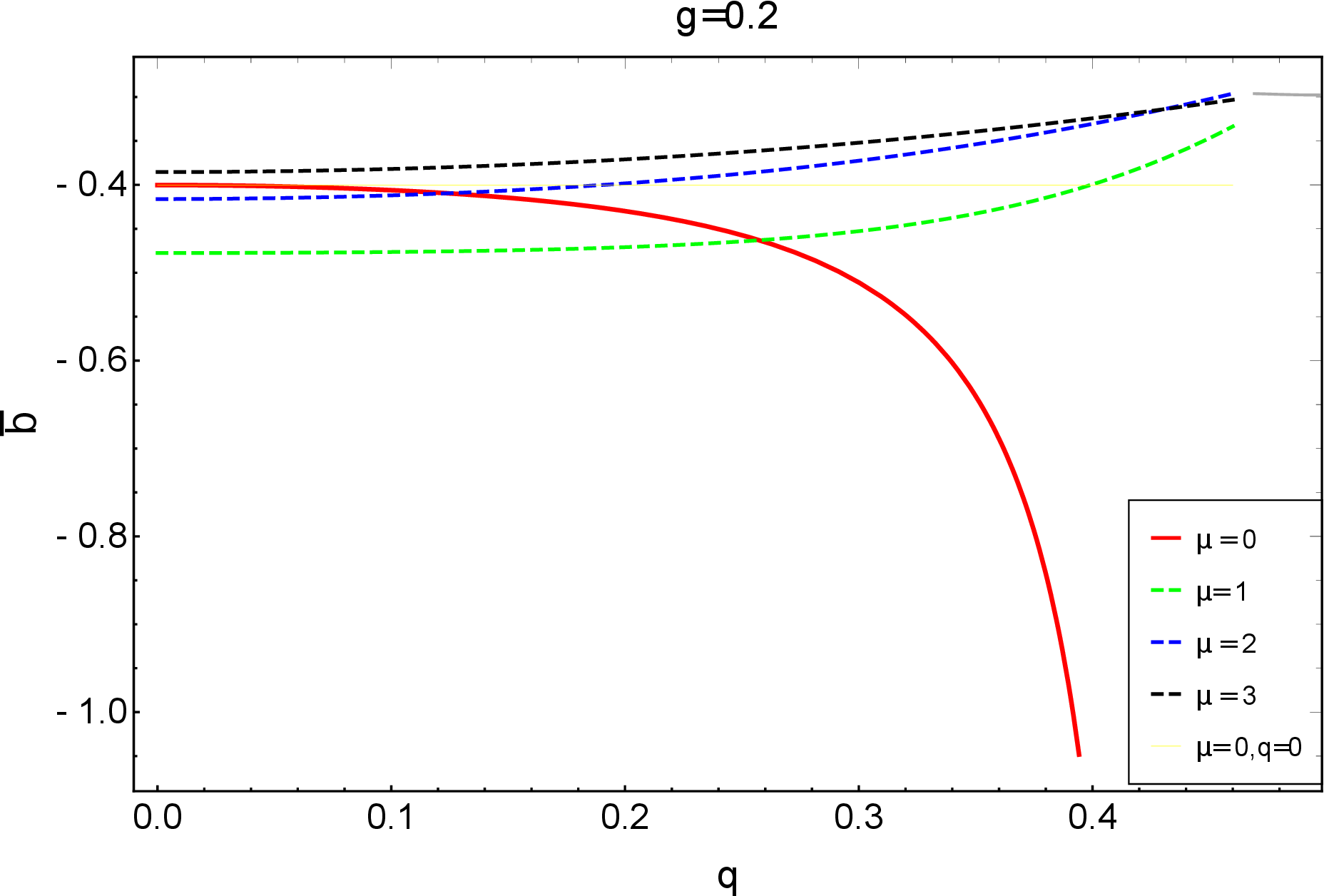}
\qquad
\includegraphics[width=.45\textwidth]{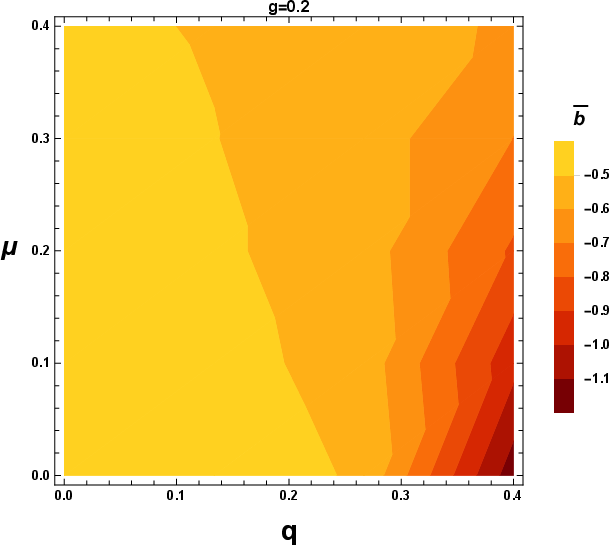}
  \caption{The behaviour of the   deflection limit coefficient  $\mathit{\bar{b}}$ vs  parameter $q$ with the different values of $\mu$ for fixed value of $g=0.2$ (left panel) ;and the   deflection limit coefficient  $\mathit{\bar{b}}$ as a function of both the parameters  $\mu$ and $q$ for the fixed value of $g=0.2$ (right panel) .}
    \label{fig:6}
\end{figure*}

\begin{table*}
 \caption{Estimation  of strong lensing coefficients with the different value of black hole parameters  $\mu=0,1,3$ ;$g=0.2,1.2$; and $|q|=0,0.05,0.1,0.2,0.4$ .\label{table:2}}
\begin{tabular}{p{2.5cm}| p{1.5cm} p{3.5cm}| p{3.5cm}p{3.5cm}| p{1.5cm} }
\hline
\hline
\multicolumn{6}{c}{Strong Lensing Coefficients}\\
$\mu$ & $g$ &$|q|$ & $\mathbf{\bar{a}}$ & $\mathbf{\bar{b}}$&$\mathbf{u_{ph}/R_{sh}}$ \\
 \tableline 
0 & ~ &$0$&1 &$-0.40023$&$2.59808$\\
\tableline 
~& ~ &$0.05$&1.0028 &$-0.401611$&$2.59808$\\
0 & ~ &$0.1$&1.101151 &$-0.406055$&$2.59373$  \\
~ & ~ &$0.2$&1.05179 &$-0.429858$&$2.58054$\\
~ & ~ &$0.4$&1.55181 &$-1.18175$&$2.22043$\\
\tableline 
~ & 0.2 &$0.05$&0.921272 &$-0.477287$&$2.955$\\
~ & 0.2 &$0.1$&0.923386 &$-0.476401$&$2.94772$  \\
~ & 0.2 &$0.2$&0.931409 &$-0.470916$&$2.91837$\\
1 & 0.2 &$0.4$&0.949693 &$-0.399493$&$2.79827$\\
\\
~ & 1.2 &$0.05$&1.03705 &$-0.516087$&$2.78908$\\
~ &1.2 &$0.1$&1.04578 &$-0.52368$&$2.77813$  \\
~ & 1.2 &$0.2$&1.08555 &$-0.561458$&$2.73253$\\
~ & 1.2 &$0.4$&1.46926 &$-1.19512$&$2.50259$\\
\tableline 
~ & 0.2 &$0.05$&0.761981 &$-0.384548$&$3.5052$\\
~ & 0.2 &$0.1$&0.761421 &$-0.381893$&$3.50177$  \\
~ & 0.2 &$0.2$&0.75901 &$-0.37101$&$3.48813$\\
3 & 0.2 &$0.4$&0.746808 &$-0.324242$&$3.435$\\
\\
~ & 1.2 &$0.05$&0.986331 &$-0.641022$&$3.21394$\\
~ &1.2 &$0.1$&0.989931 &$-0.644732$&$3.20768$  \\
~ & 1.2 &$0.2$&1.00474 &$-0.660005$&$3.18241$\\
~ & 1.2 &$0.4$&1.06929 &$-0.7222588$&$3.07755$\\
\hline 
\hline 
\end{tabular}
\end{table*}

\begin{figure*}[htbp]
\centering
\includegraphics[width=.45\textwidth]{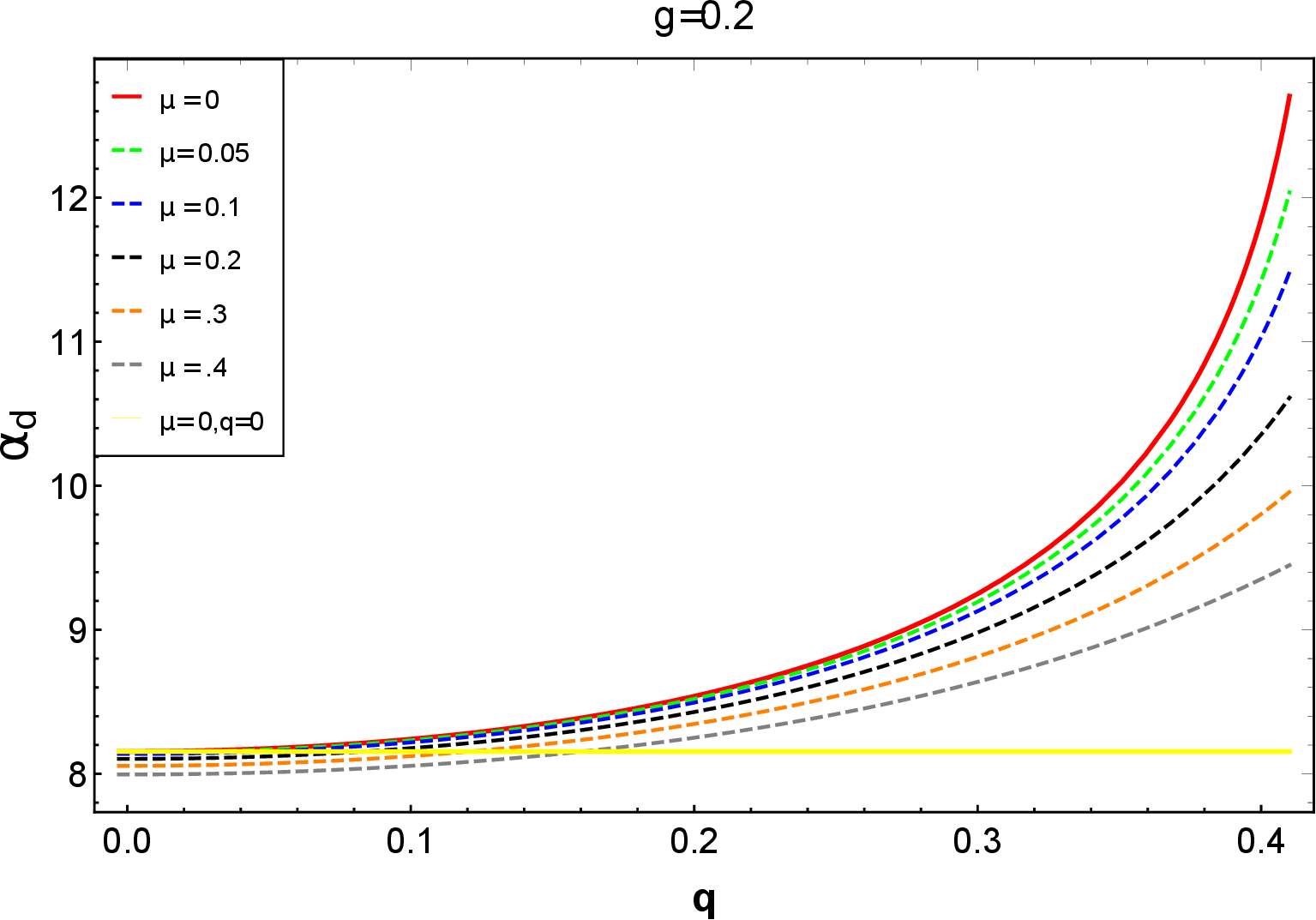}(a)
\qquad
\includegraphics[width=.45\textwidth]{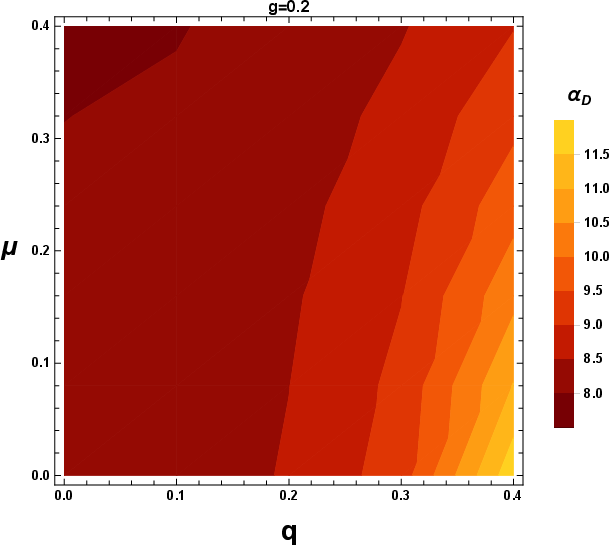}(b)
\vfill
\vfill
\centering
\includegraphics[width=.4\textwidth]{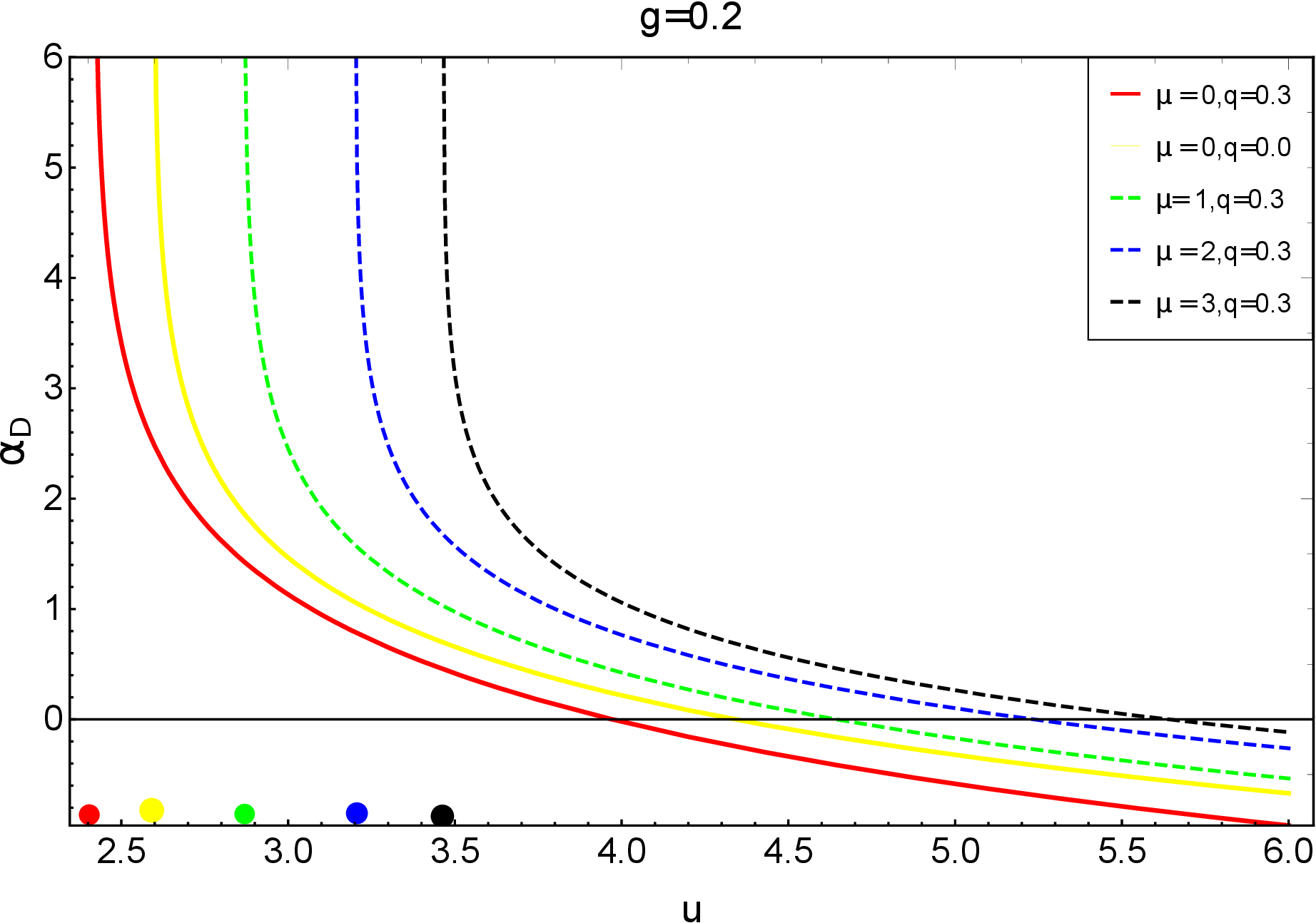}(c)
\qquad
\includegraphics[width=.4\textwidth]{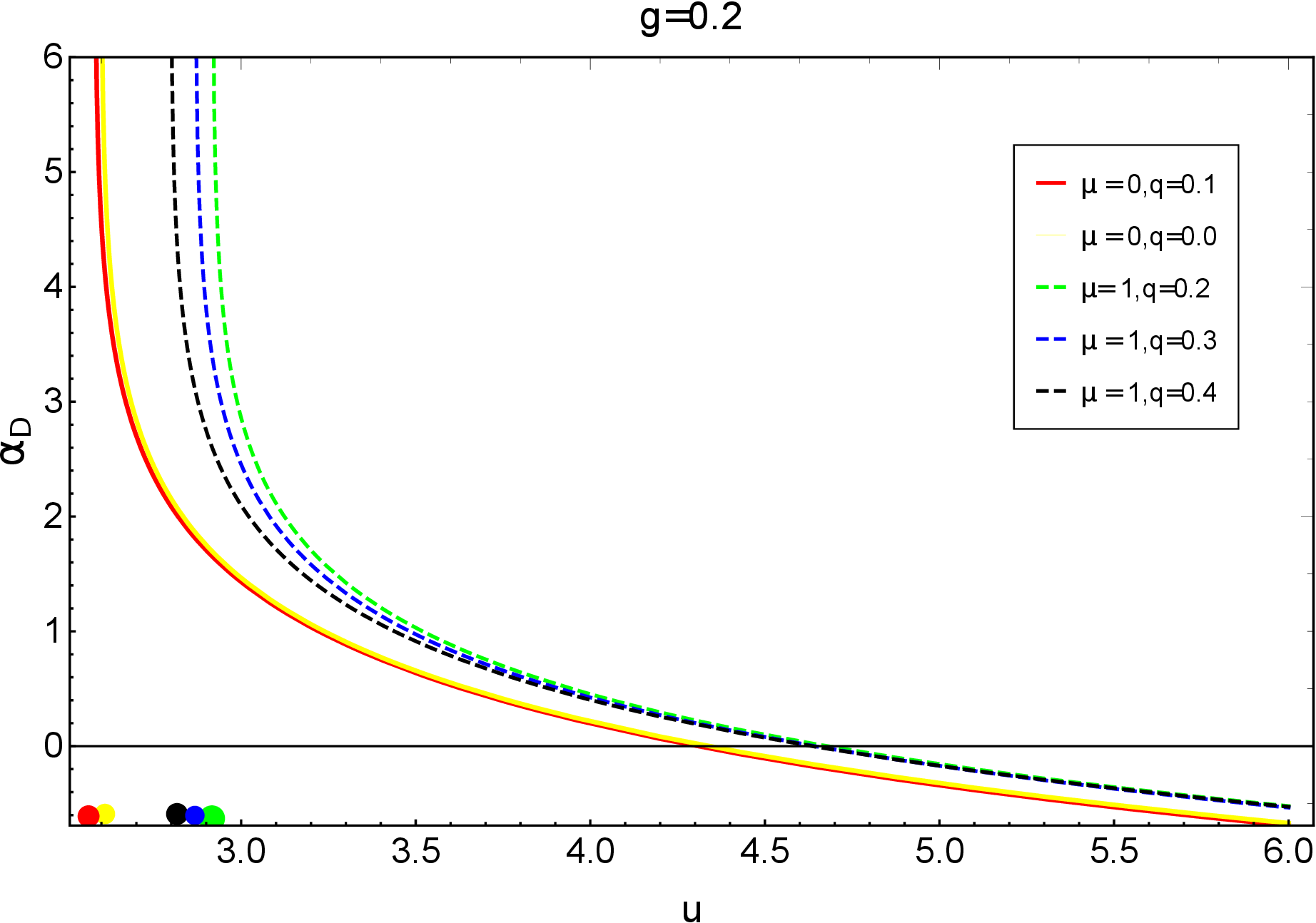}(d)
\caption{The behaviour of the deflection angle $\mathit{\alpha_D}$ vs  parameter $q$ with the different values of $\mu$ for fixed value of $g=0.2$ ( panel (a)) ; the   deflection angle $\mathit{\alpha_D}$ as a function of both the parameters  $\mu$ and $q$ for the fixed value of $g=0.2$ (panel (b)) ;the   deflection angle $\mathit{\alpha_D}$ as a function of impact parameter $u$  with different value of $q$ for the fixed value of parameters  $\mu$ and $q$ (panel (c)); and the   deflection angle $\mathit{\alpha_D}$ as a function of impact parameter $u$  with different value of $\mu$ for the fixed value of parameters  $\mu$ and $q$ (panel (d)).The solid red line corresponds to the case of regular Bardeen and the solid yellow line corresponds to the case of modified  Bardeen black hole. The dots in panels (c)\&(d) indicate the values of the impact parameter $u=u_{ph}$, where the deflection angle $\mathit{\alpha_D}$ becomes divergent.  }
    \label{fig:7}
\end{figure*}

We've obtained the numerical values of the lensing coefficients \(\bar{a}\) and \(\bar{b}\) alongside \(u_{ph}/R_{sh}\) for various parameters of the modified Bardeen black hole: \(\mu=0,1,3\), \(g=0.2,1.2\), and \(q=0,0.05,0.1,0.2,0.4\) (refer to Table.\ref{table:2}). Observations from this table indicate that, for fixed parameters \(g=0.2,1.2\) and \(\mu=0,1,3\), the lensing coefficient \(\bar{a}\) increases as the parameter \(q\) grows in magnitude, while the lensing coefficient \(\bar{b}\) decreases. However, this behavior doesn't hold for the case when \(g=0.2\) and \(\mu=3\). Specifically, when \(\mu=0\) and \(q=0\), the values of the lensing coefficients are \(\bar{a}=1\) and \(\bar{b}=-0.40023\), aligning with the characteristics of the Schwarzschild black hole \cite{Bozza:2002zj}.

The behaviors of the lensing coefficients \(\bar{a}\) and \(\bar{b}\) are visually displayed in Figs.\ref{fig:5} and \ref{fig:6}, respectively.

Moreover, the behavior of the deflection angle \(\alpha_D\) around the modified Bardeen black hole is presented in Fig.\ref{fig:7}. Observations from Figs.\ref{fig:7}(a) and \ref{fig:7}(b) indicate that \(\alpha_D\) increases with the increasing magnitude of the charge parameter \(q\) for small values of \(\mu\) while keeping \(g\) fixed. Conversely, \(\alpha_D\) decreases with an increasing value of \(\mu\), assuming other parameters remain constant. Furthermore, the deflection angle \(\alpha_D\) for the modified Bardeen black hole lies between the values for the Schwarzschild black hole (\(\mu=0, q=0\)) and the ordinary regular Bardeen black hole (\(\mu=0\)).

Lastly, in Fig.\ref{fig:7}(c), \(\alpha_D\) decreases with the critical impact parameter \(u\) for various values of the parameter \(\mu\) while \(q\) and \(g\) are fixed. In Fig.\ref{fig:7}(d), \(\alpha_D\) decreases with different magnitude values of parameter \(q\) for fixed values of \(\mu\) and \(g\).

\subsection{Lensing observables}

We now go through the analysis of strong lensing effects caused by the modified Bardeen black hole. In this investigation, we consider a scenario where both the observer and the source are positioned at a significant distance from the black hole (acting as the lens) and are nearly in alignment. Additionally, we assume that the source lies behind the black hole. Consequently, we can establish the lens equation \cite{Bozza:2001xd} as:

\begin{equation}\label{24}
 \beta=\theta-\frac{D_{ls}}{D_{os}}\Delta \alpha_{n}
\end{equation}

Here, \(\Delta\alpha_{n}=\alpha_D(\theta) - 2n\pi \) denotes the offset deflection angle, with \(n\) indicating the number of loops made by the photon ray around the black hole. In this context, \(\beta\) and \(\theta\) represent the angular separations respectively between the black hole (lens) and the source, and between the observer and the source. Furthermore, \(D_{ls}\), \(D_{ol}\), and \(D_{os}\) stand for the distances of lens-source, observer-lens, and observer-source, satisfying the condition \(D_{os}=D_{ol}+D_{ls}\).

Using the Eqns. (\ref{19}) and (\ref{24}), the angular separation between the  black hole(lens) to the  $n^{th}$ relativistic  image can be
expressed as
\begin{equation}\label{25}
     \theta _n =  \theta^0 _n - \frac{u_{ph}e_n(\theta_n^0-\beta)D_{os}}{\bar{a}D_{ol}D_{ls}}
 \end{equation}
where $$ e_n=e^{\frac{\bar{b}-2n\pi}{\bar{a}}},$$
$$\theta^0_n=\frac{u_{ph}(1+e_n)}{D_{ol}}$$.
In the scenario where a photon completes \(2n\pi\) loops around the black hole (lens), \(\theta^0_n\) denotes the angular position of the resultant image.

In the context of strong gravitational lensing, where surface brightness is conserved, the magnification of the \(n^{th}\) relativistic image is defined as the ratio between the solid angle subtended by this specific image and that of the source \cite{Virbhadra:1999nm}. The expression for the magnification of the \(n^{th}\) relativistic image is derived as stated in \cite{Bozza:2002zj}.
\begin{equation}\label{26}
\mu_n=\biggr(\frac{\beta}{\alpha}\frac{d\beta}{d\alpha}\biggr)^{-1}\biggr|_{\theta_0}=\frac{ u^2_{ph}(1+e_n)e_n D_{os}}{\beta \bar{a}D_{ls}D^2_{ol}}
\end{equation}

The equation above implies that the initial relativistic image appears as the brightest, and its brightness diminishes exponentially with each subsequent \(n\). Consequently, the luminosity of this image predominates over the others. Equation (\ref{26}) exhibits divergence when \(\beta \rightarrow 0\), indicating that perfect alignment maximizes the probability of detecting relativistic images.

Consideration is given to the scenario where the brightest image, i.e., the outermost image \(\theta_1\), is observed as a solitary image, while the remaining inner images coalesce at \(\theta_{\infty}\) (\(\theta_n|_{n\rightarrow \infty}=: \theta_{\infty}\)). Leveraging the deflection angle in equation (\ref{21}), we can derive essential strong lensing parameters. These include the angular position of the image set \(\theta_{\infty}\), the angular separation between the outermost and innermost images \(S\), and the relative magnification \(r_{mag}\) between the outermost relativistic image and the grouped inner relativistic images \cite{Kumar:2022fqo,Bozza:2002zj}.

\begin{equation}\label{27}
\theta_{\infty}=\frac{u_{ph}}{d_{ol}}
\end{equation}
 \begin{equation}\label{28}
S= \theta_1-\theta_{\infty}\approx\theta_{\infty}e^\frac{(\bar{b} -2\pi)}{\bar{a}}
  \end{equation}

  \begin{equation}\label{29}
  r_{mag}=\frac{\mu_1}{\Sigma^\infty_{n=2}\mu_{n}}\approx \frac{5\pi}{\bar{a}log(10)}
  \end{equation}

Upon measurement of the strong lensing observables—\(\theta_{\infty}\), \(S\), and \(r_{mag}\)—from observations, the lensing coefficients \(\bar{a}\), \(\bar{b}\), and the minimum impact parameter \(u_{ph}\) can be readily derived by inversely solving Eqns. (\ref{27}), (\ref{28}), and (\ref{29}). These obtained values can then be compared with the theoretically calculated ones. Such a comparative analysis allows for the identification and differentiation among the modified Bardeen, ordinary regular Bardeen, and Schwarzschild black holes based on these observations.\\

Taking into account the supermassive black holes $M87^*$, $Sgr A^*$, and $NGC 7457$ found in nearby galaxies, we proceed to estimate the observable quantities—\(\theta_{\infty}\), \(S\), and \(r_{mag}\)—within the framework of a modified Bardeen black hole (Refer to Table 4). The respective mass and distances from Earth for these black holes are as follows: For $M87^*$ \cite{EventHorizonTelescope:2019pgp,EventHorizonTelescope:2019ggy}, the values are approximately \(M \approx 6.5 \times 10^9 M_{\odot}\) and \(D_{ol} \approx 16.8\) Mpc; for $Sgr A^*$, the values are approximately \(M \approx 4.28 \times 10^6 M_{\odot}\) and \(D_{ol} \approx 8.32\) kpc \cite{EventHorizonTelescope:2022wkp,EventHorizonTelescope:2022apq}; and for $NGC 7457$, the values are approximately \(M \approx 8.95 \times 10^6 M_{\odot}\) and \(D_{ol} \approx 12.53\) Mpc \cite{Kormendy:2013dxa}.

\begin{figure*}[htbp]
\centering
\includegraphics[width=.45\textwidth]{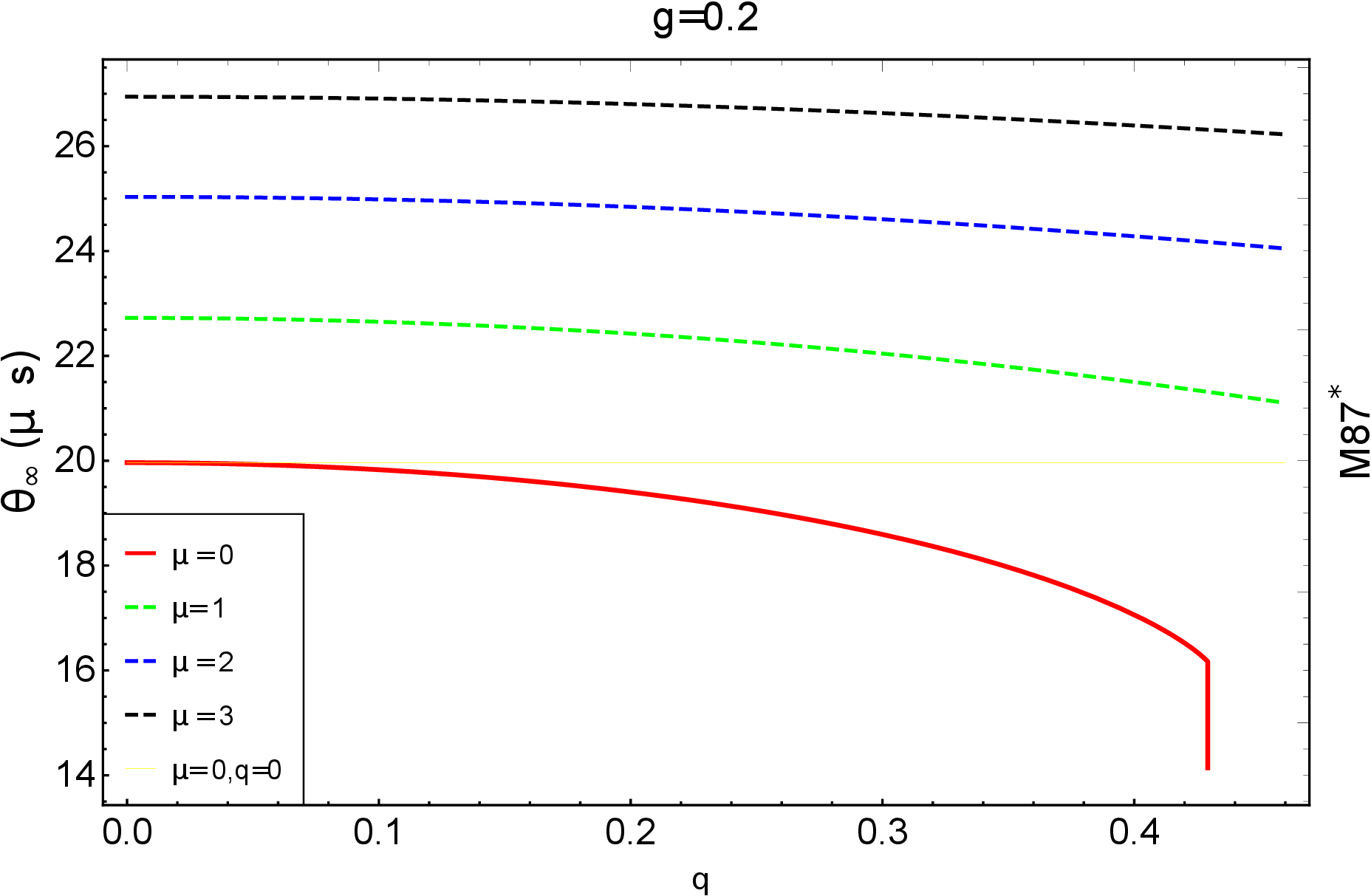}(a)
\qquad
\includegraphics[width=.45\textwidth]{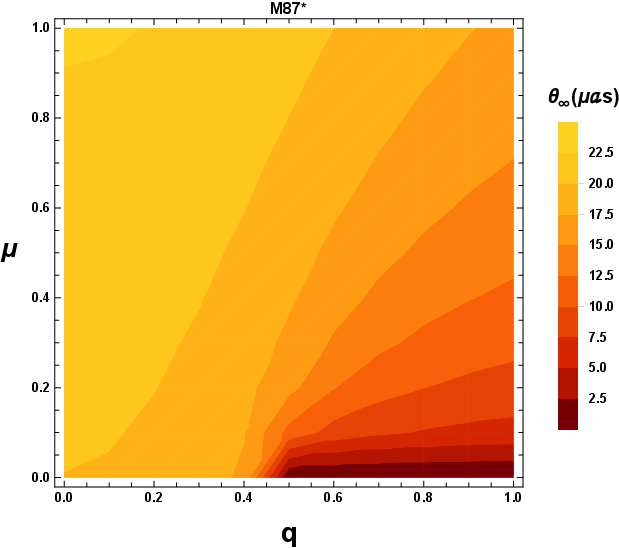}(b)
\vfill
\vfill
\centering
\includegraphics[width=.45\textwidth]{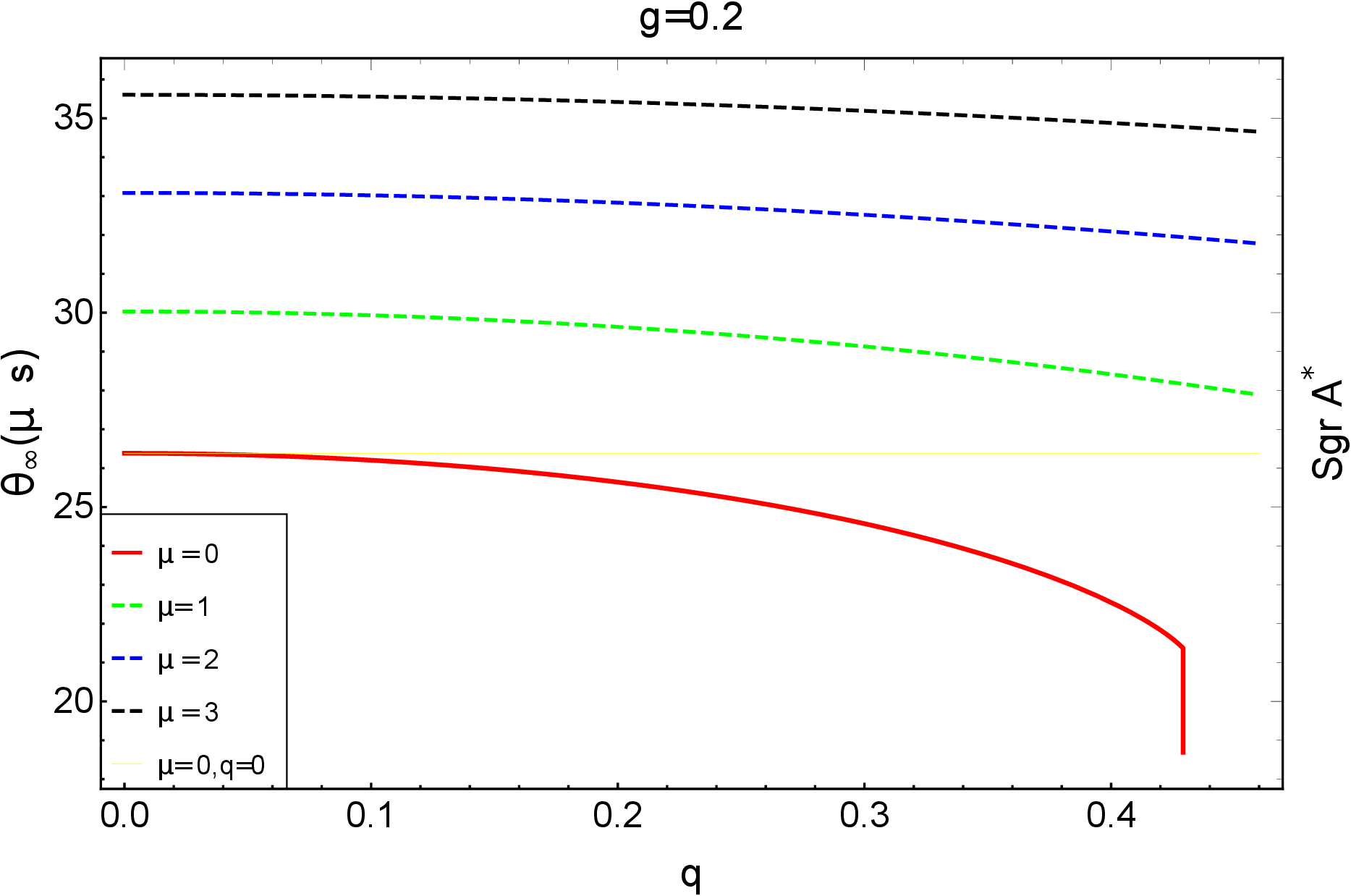}(c)
\qquad
\includegraphics[width=.45\textwidth]{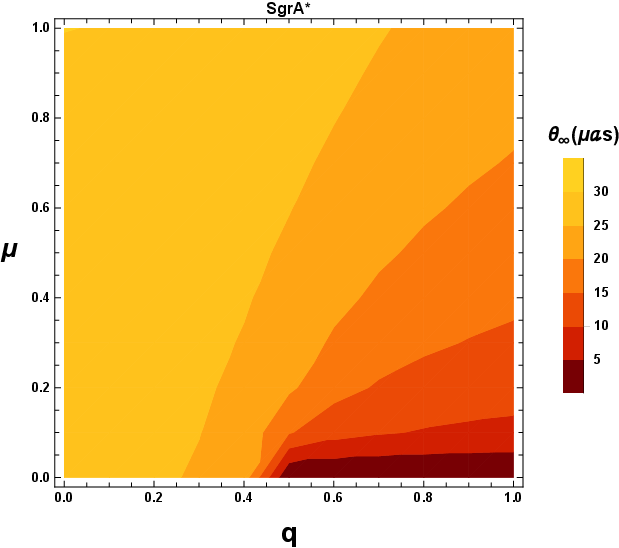}(d)
 \caption{The behaviour of the  angular image position $\mathit{\theta_{\infty}}$ vs  parameter $q$ with the different values of $\mu$ for the fixed value of $g=0.2$ for $M87^{*}$(upper left panel)  and for $SgrA^{*}$ (lower left panel)   ;and the  angular image position    $\mathit{\theta_{\infty}}$ as a function of both the parameters  $\mu$ and $q$ for the fixed value of $g=0.2$  for $M87^{*}$ (lower right panel) and for $SgrA^{*}$ (lower right panel).}
    \label{fig:8}
\end{figure*}

\begin{figure*}[htbp]
\centering
\includegraphics[width=.45\textwidth]{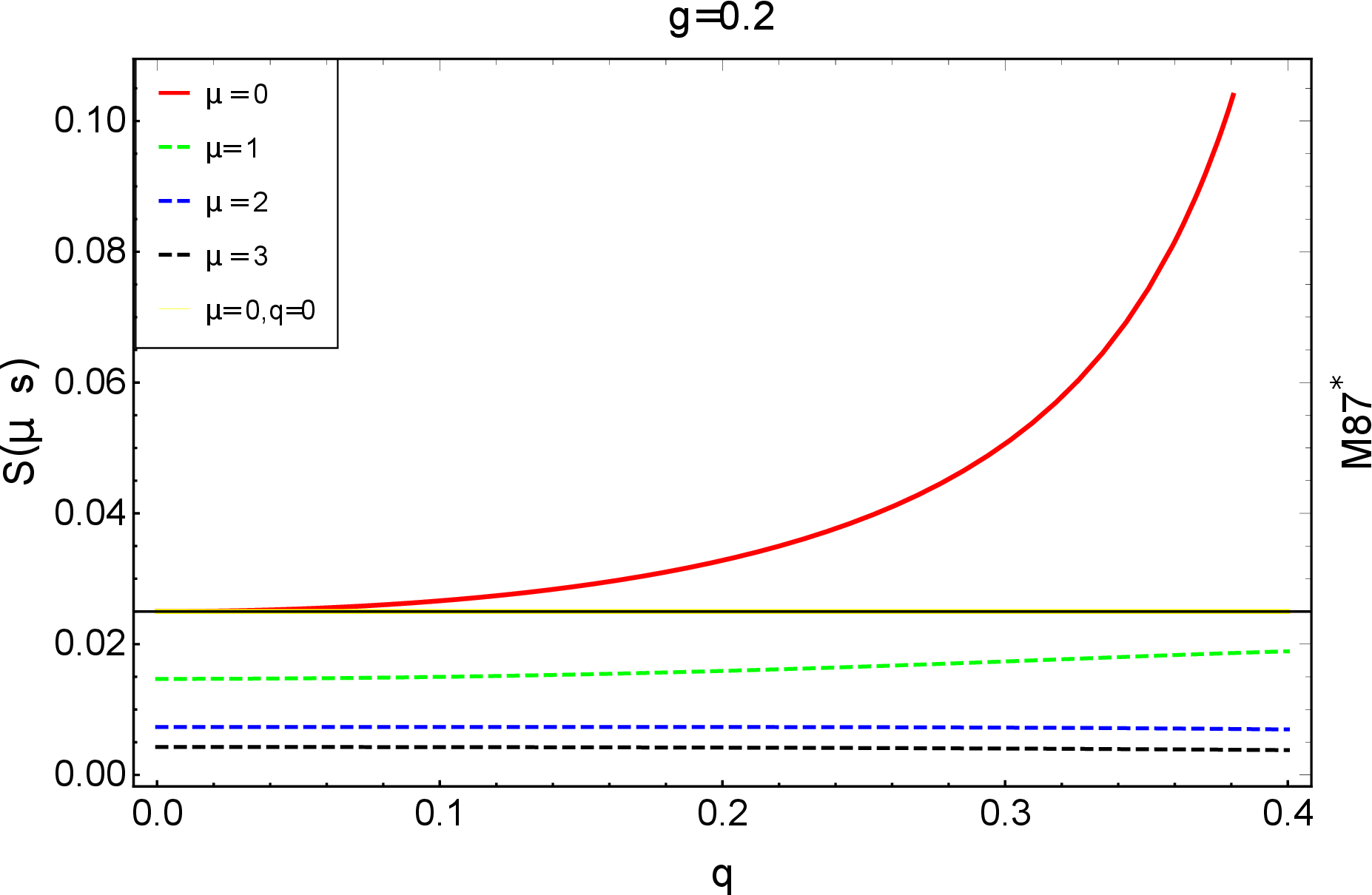}(a)
\qquad
\includegraphics[width=.45\textwidth]{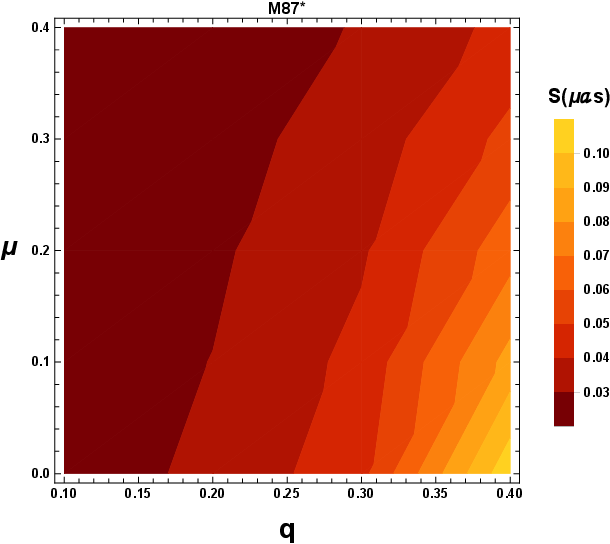}(b)
\vfill
\vfill
\centering
\includegraphics[width=.45\textwidth]{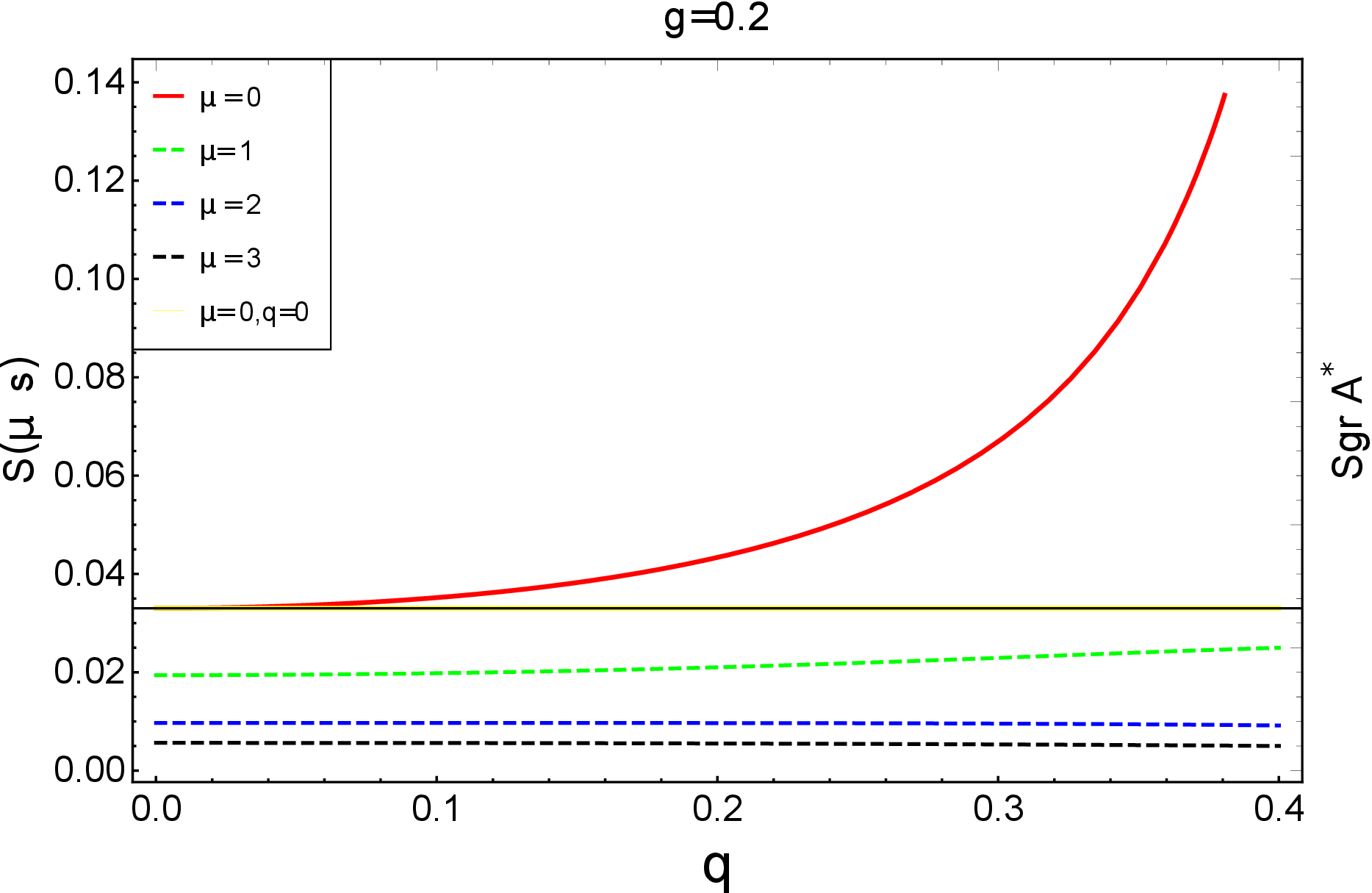}(c)
\qquad
\includegraphics[width=.45\textwidth]{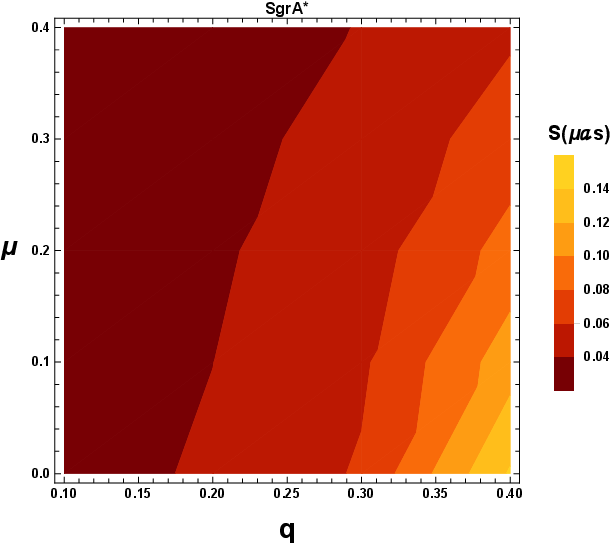}(d)
 \caption{The behaviour of the  angular image separation $\mathit{S}$ vs  parameter $q$ with the different values of $\mu$ for the fixed value of $g=0.2$ for $M87^{*}$(upper left panel)  and for $SgrA^{*}$ (lower left panel)   ;and the  angular image separation $\mathit{S}$  as a function of both the parameters  $\mu$ and $q$ for the fixed value of $g=0.2$  for $M87^{*}$ (lower right panel) and for $SgrA^{*}$ (lower right panel).}
    \label{fig:9}
\end{figure*}

\begin{figure*}[htbp]
\centering
\includegraphics[width=.45\textwidth]{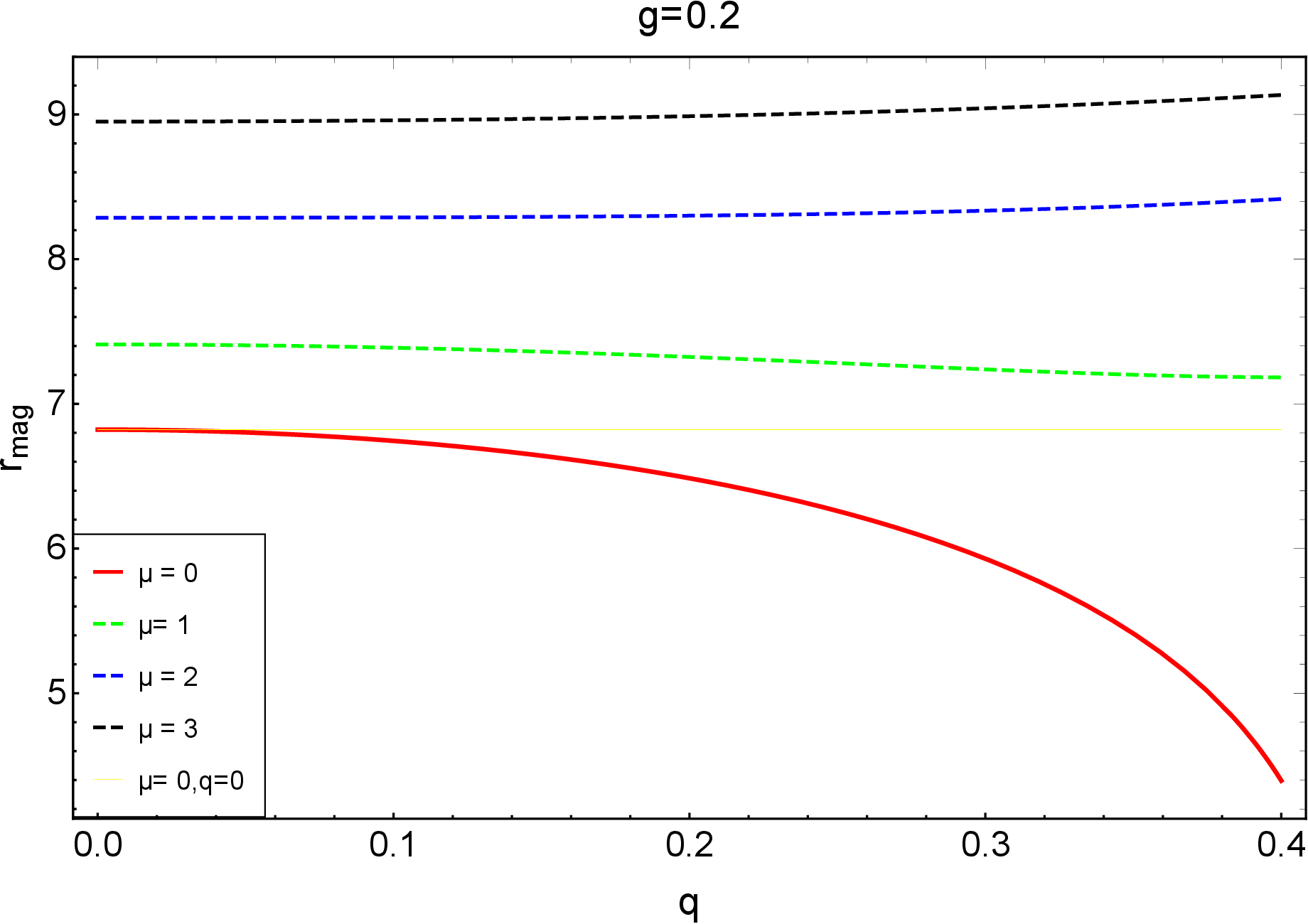}
\qquad
\includegraphics[width=.45\textwidth]{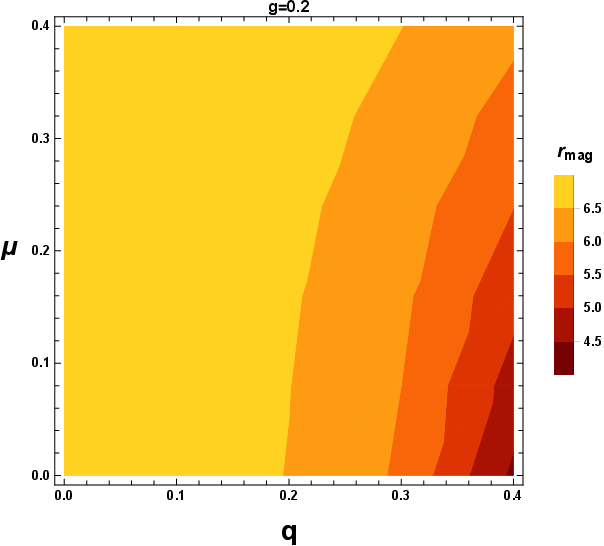}
 \caption{The behaviour of the relative magnification $\mathit{r_{mag}}$ vs  parameter $q$ with the different values of $\mu$ for the fixed value of $g=0.2$ (left panel) ;and the relative magnification $\mathit{r_{mag}}$ as a function of both the parameters  $\mu$ and $q$ for the fixed value of $g=0.2$ (right panel) .Note that the relative magnification $\mathit{r_{mag}}$ does not the mass or distance of the black hole.}
    \label{fig:10}
\end{figure*}

 \begin{table*}
\begin{center}
 \caption{Estimation  of strong lensing observables for supermassive BHs $ M87^*$,$ Sgr A^*$, $NGC 7457$ with the different value of black hole parameters  $\mu=0,1,3$ ;$g=0.2,1.2$; and $|q|=0,0.05,0.1,0.2,0.4$.The observable quantity $r_{mag}$ does not depend on the mass or distance of the black hole from the observer.\label{table:3}}
 \begin{tabular}{ccc|cc|cc|cc|c}
 \hline
 \hline
&{parameters} && {$ M87^*$}& &{$ Sgr A^*$ }&&{ $NGC 7457$}&&$ M87^*$,$ Sgr A^*$ , $NGC 7457$
\\
$\mu$& $g$ & $|q|$ & $ \theta_{\infty} (\mu as)$&$S(\mu as)$&
$\theta_{\infty} (\mu as)$ &  $S (\mu as)$ & $\theta_{\infty} (\mu as)$ &$S(\mu as)$&$r_{mag}$ \\

\hline
\hline
{0} & 0 & 0&19.9633&0.024984&26.3315&0.0329538&0.0365211&$4.57\times 10^{-5}$&6.82188\\
\hline
&  & 0.05 &19.9298&0.025377&26.2874&0.0334722&0.03646&$4.63\times 10^{-5}$&6.80283\\
0&  & 0.1 &19.8285&0.0266224&26.1537&0.0351149&0.0362745&$4.87\times 10^{-5}$&6.74426\\
& & 0.2 &19.402&0.0328069&025.5912&0.0432723&0.0354944&$6.0\times 10^{-5}$&6.48597\\
&  & 0.4 &17.0615&0.138949&22.504&0.183274&0.0312125&$2.54\times 10^{-4}$&4.39608\\
\hline
&& 0.05 &22.7058&0.0147639&29.9489&0.0194735&0.0415383&$2.70\times 10^{-5}$&7.40485\\
&0.2 &0.1 &22.6499&0.0149914&29.8751&0.0197736&0.041436&$2.74\times 10^{-5}$&7.3879\\
& &0.2 &22.4243&0.0159015&29.5776&0.020974&0.0410234&$2.91\times 10^{-5}$&7.32426\\
1& &0.4&21.5015&0.0189008&28.3604&0.0249302&0.0393352&$3.46\times 10^{-5}$&7.18325\\
\\
&& 0.05 &21.4309&0.0304544&28.2673&0.0401692&0.039206&$5.57\times 10^{-5}$&6.57816\\
&1.2 &0.1 &21.3467&0.0318095&28.1563&0.0419567&0.0390521&$5.82\times 10^{-5}$&6.52325\\
& &0.2 &20.9964&0.0383546&27.6942&0.0505896&0.038411&$7.02\times 10^{-5}$&6.28426\\
& &0.4&19.2295&0.118434&25.3637&0.156215&0.0351788&$2.17\times 10^{-4}$&4.64307\\
\hline
&& 0.05 &26.9335&0.00426567&35.5252&0.00562641&0.0492725&$7.8\times 10^{-6}$&8.95282\\
&0.2 &0.1 &26.9071&0.00424895&35.4904&0.00560436&0.0492242&$7.77\times 10^{-6}$&8.95941\\
& &0.2 &26.8023&0.00417578&35.3522&0.00550785&0.0490325&$7.64\times 10^{-6}$&8.98787\\
3& &0.4&26.394&0.00379368&34.8137&0.00500385&0.0482857&$6.94\times 10^{-6}$&9.13472\\
\\
&& 0.05 &24.6955&0.0220698&32.5732&0.02911&0.0451782&$4.03\times 10^{-5}$&6.91642\\
&1.2 &0.1 &24.6473&0.0225118&32.5098&0.029693&0.0450902&$4.11\times 10^{-5}$&6.89127\\
& &0.2 &24.4532&0.0243877&32.2537&10.0321674&0.044735&$4.46\times 10^{-5}$&6.7897\\
& &0.4&23.6474&0.0337682&31.1909&0.0445402&0.043261&$6.17\times 10^{-5}$&6.37982\\
\hline
\hline
\end{tabular}
\end{center}
\end{table*}

The behavior of strong lensing observables—angular image position \(\theta_{\infty}\), angular image separation \(S\), and relative magnification \(r_{mag}\)—is illustrated in Figs. \ref{fig:8}, \ref{fig:9}, and \ref{fig:10}, respectively, as a function of the parameter \(q\). These figures also depict the behavior concerning both parameters \(q\) and \(\mu\) for a fixed value of \(g=0.2\) concerning \(M87^{*}\) and \(SgrA^{*}\) (refer to Table 3 as well).

It's notable that \(\theta_{\infty}\) and \(r_{mag}\) exhibit a decrease with an increase in the magnitude of the charge parameter \(q\), while \(S\) displays an increment with the rising magnitudes of both parameters \(\mu\) and \(q\), given other parameters remain fixed. Moreover, \(r_{mag}\) increases with higher values of parameter \(\mu\) for a constant \(g\) and \(q\). Additionally, the relative magnification \(r_{mag}\) for the modified Bardeen black hole surpasses that of both the Schwarzschild (\(\mu=0, q=0\)) and ordinary regular Bardeen (\(\mu=0\)) black holes.

\subsection{Einstein Ring}

When the source, black hole (lens), and observer are perfectly aligned
i.e., when $\beta=0$, a black hole (lens) deflects the light rays  in all
direction  such that a ring-shaped image is produced, which is called an Einstein ring
\cite{Einstein:1936llh,Liebes:1964zz,Mellier:1998pk,Bartelmann:1999yn,Schmidt:2008hc,Guzik:2009cm}.

By simplifying the Eqn. \ref{25} for $\beta=0$,we obtain the angular radius of $n^{th}$ relativistic images as follows:
\begin{equation}\label{30}
     \theta _n =  \theta^0 _n \biggr(1 - \frac{u_{ph} e_n  D_{os}}{\bar{a}D_{ls}D_{ol}}\biggr)
 \end{equation}

Considering  the case where the black hole (lens) is  located at a half distance between the source
and receiver i.e., $D_{os}=2D_{ol}$ and taking $D_{ol}>>u_{ph}$, thus the
angular radius of the $n^{th}$ relativistic Einstein ring in the context of a modified Bardeen black hole is given by

\begin{equation}\label{31}
\theta^E_n=\frac{ u_{ph}(1+e_n)}{D_{ol}}
\end{equation}

\begin{figure*}[htbp]
\centering
\includegraphics[width=.45\textwidth]{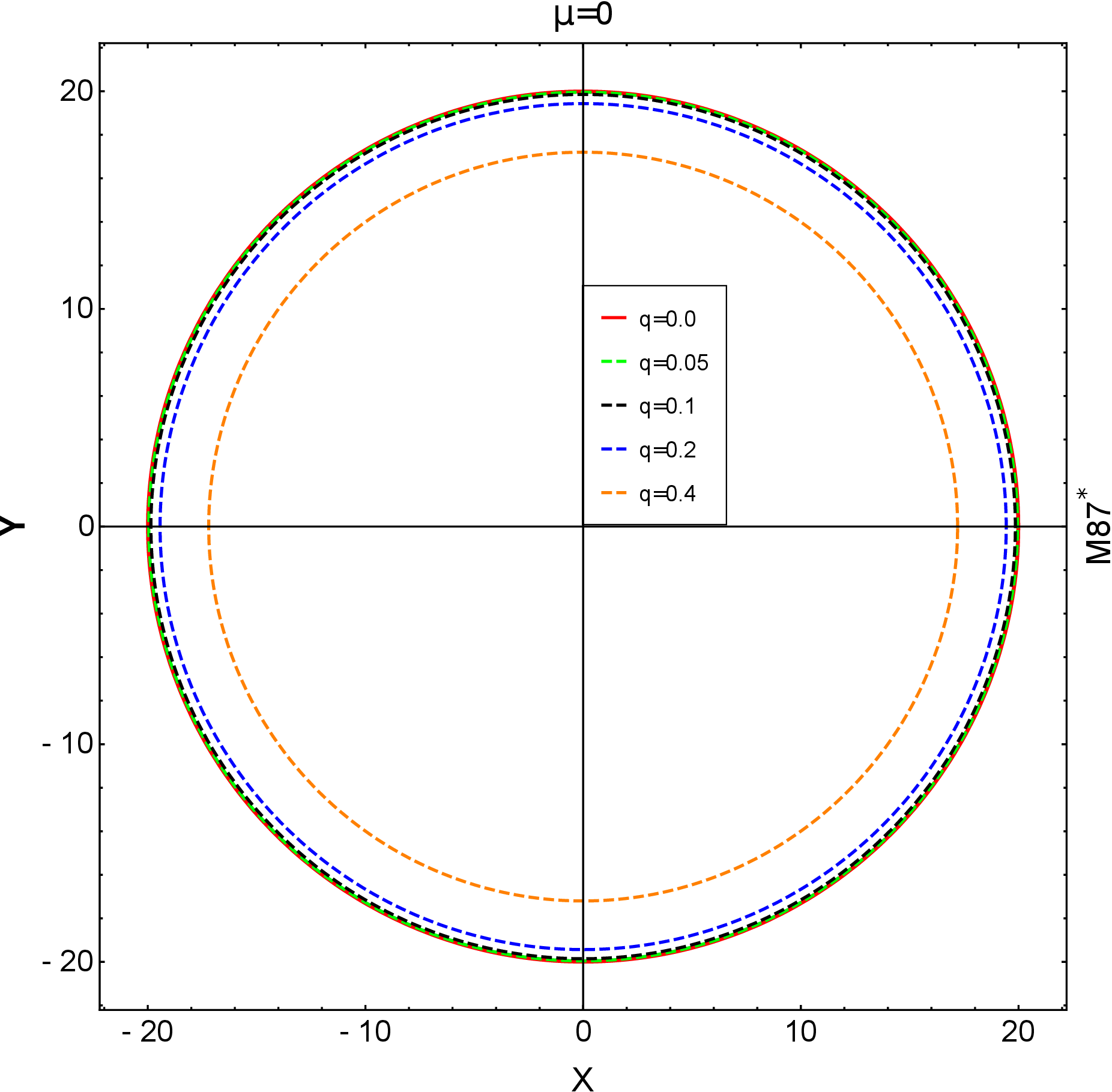}(a)
\qquad
\includegraphics[width=.45\textwidth]{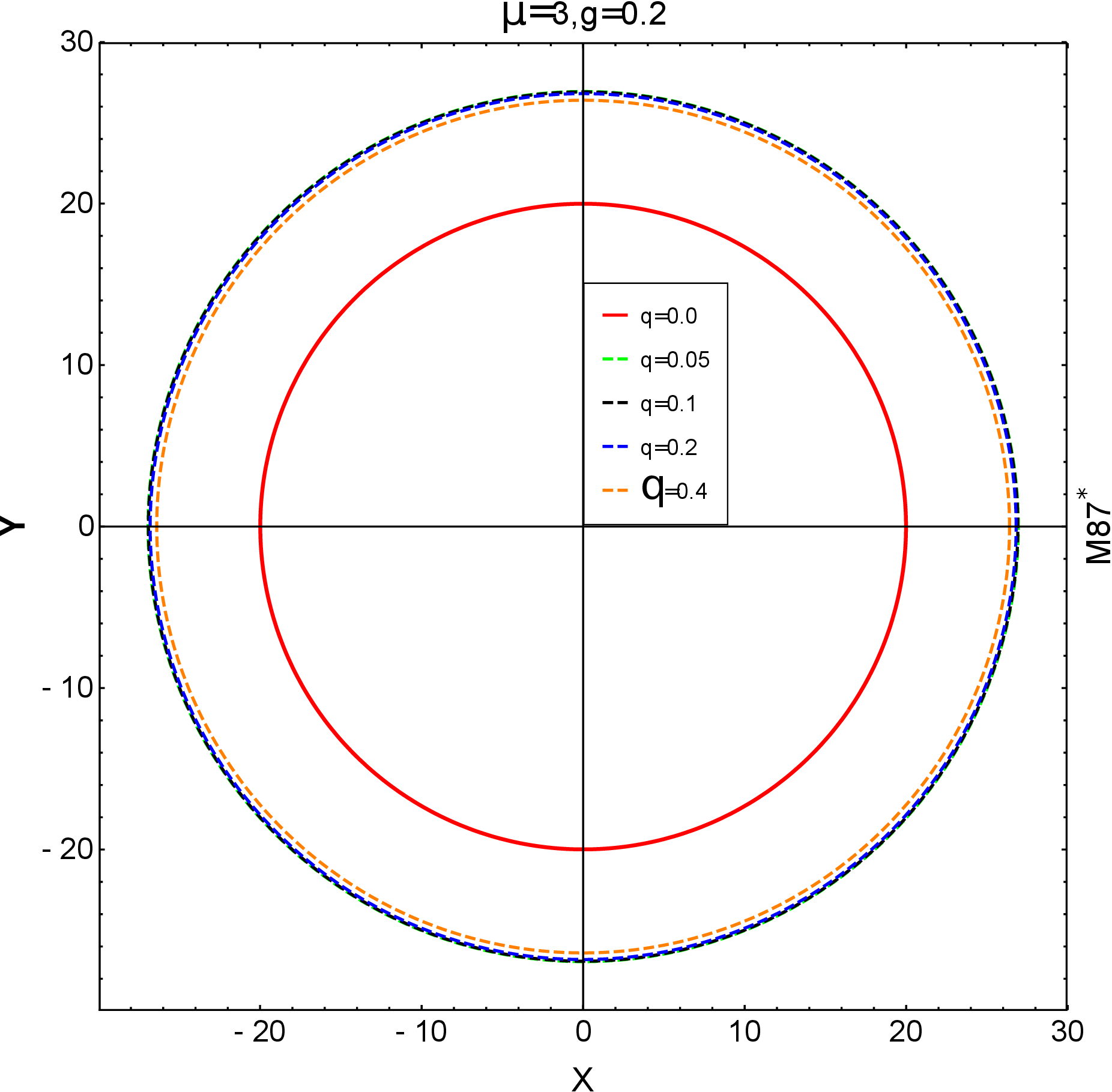}(b)
\vfill
\vfill
\centering
\includegraphics[width=.45\textwidth]{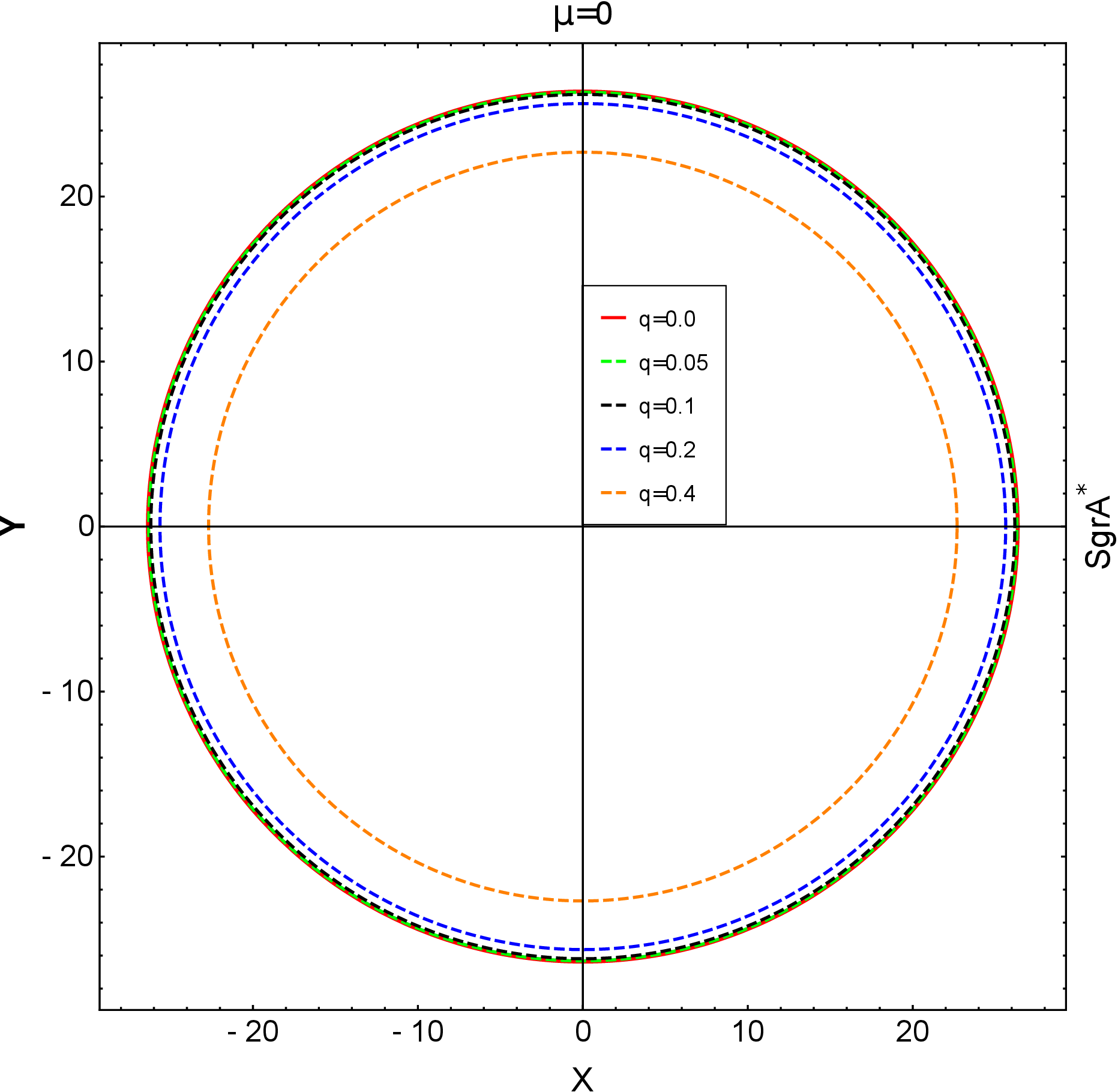}(c)
\qquad
\includegraphics[width=.45\textwidth]{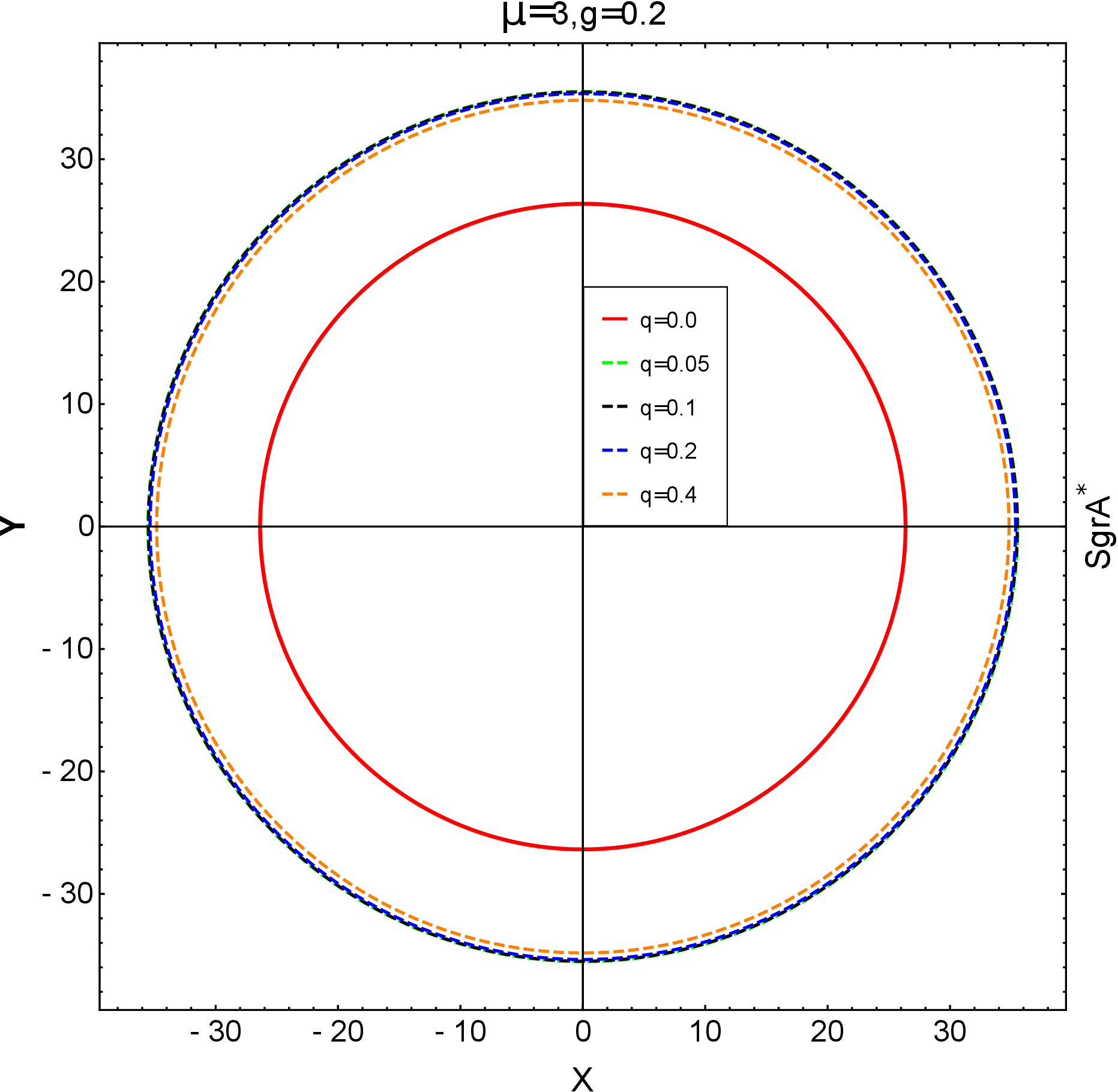}(d)
 \caption{The behavior of the outermost Einstein's ring $\mathit{\theta^{E}_{n}}$  for the case of regular Bardeen  (panel a \& c) for $M87^{*}$ and  $SgrA^{*}$; and the case of modified  Bardeen black hole  (panel b \& d) for $M87^{*}$ and  $SgrA^{*}$ respectively.The solid circular ring corresponds to the case of Schwarzschild's black hole. }
    \label{fig:11}
\end{figure*}

The angular radius ($\theta^E_1$) represents the outermost Einstein ring, as depicted in Fig.\ref{fig:11} for the supermassive black holes ( $M87^*$ ) (Fig. \ref{fig:11}(a) \& (b)) and ( $Sgr A^*$ ) (Fig. \ref{fig:11}(c) \& (d)).

It's evident that under fixed parameters \(\mu\) and \(g\), the outermost Einstein rings exhibit a decrease with an increasing magnitude of the parameter \(q\) concerning both the supermassive black holes \( Sgr A^* \) and \( M87^* \). Additionally, it's noteworthy that the outermost Einstein rings for the modified Bardeen black hole exceed those of the ordinary regular Bardeen black hole.

\subsection{Time delay in strong field Limit}

Time delay is one of the most important observable by the strong gravitational lensing phenomenon, which is obtained by the time difference between the formation of two relativistic images. The time difference is caused when the photon travels in a different path around the black hole. The time travel by the different photon paths for the different relativistic images is different and hence, there is a time difference between the different relativistic images. If the time signals of two relativistic images are distinguished from the observation, one can calculate the time delay between two signals \cite{Bozza:2003cp}. The time taken by a photon to revolve  around the black hole \cite{Bozza:2003cp} is read as
 \begin{equation}\label{32}
\tilde{T}=\tilde{a}log\biggr(\frac{u}{u_{ph}}-1\biggr)+\tilde{b}+\mathcal{O}(u-u_{ph})
\end{equation}

With the help of the above  Eq.(\ref{32}), one can compute the time difference between two relativistic images.

For spherically static symmetric black hole spacetime, the time delay between  two relativistic
images, when the relativistic images are on the same side of the black hole, are obtained as
\begin{equation}\label{33}
\Delta T_{2,1}=2\pi u_c=2\pi D_{ol} \theta_{\infty}
\end{equation}
If the time delay \(\Delta T_{2,1}\) between two relativistic images is obtained with an accuracy of \(5\%\) and the critical impact parameter \(u_{ph}\) with a negligible error is measured, it becomes feasible to determine the black hole distance with an accuracy of \(5\%\).

Numerical estimations of the time delay \(\Delta T_{2,1}\) for various supermassive black holes in the context of the standard Schwarzschild (\(\mu=0\), \(q=0\)), ordinary regular Bardeen (\(\mu =0\), \(q=0.3\)), and modified Bardeen (\(\mu =3\), \(q=0.3\)) black holes have been conducted (refer to Table \ref{table:4}).

The findings reveal that the time delay \(\Delta T_{2,1}\) between two relativistic images associated with the modified Bardeen black hole (\(\mu =3\), \(g=0.2\), \(q=0.3\)) is significantly greater compared to the cases of the standard Schwarzschild (\(\mu=0\), \(q=0\)) and ordinary regular Bardeen (\(\mu=0\), \(q=0.3\)) black holes.

\begin{table*}
 \caption{Estimation of time delay for some supermassive BHs in the context of Schwarzschild  ($\mu=0$,$q=0 $), ordinary regular Bardeen ($q=0.3,\mu=0$) and modified Bardeen ($\mu=3$,$g=0.2$,$q=0.3 $) black hole spacetimes. Mass(M) and distance $D_{ol}$
respectively are taken in solar mass and Mpc units \cite{Kormendy:2013dxa}.  Time delays $\Delta T_{2,1}$ are estimated in minutes.
\label{table:4}}
\begin{tabular}{p{2.cm}| p{2cm} |p{2.2cm}| p{2.5cm}|p{3.5cm}| p{4.2cm} }
\hline
\hline
\multicolumn{6}{c}{Time delays}\\
\hline

$ Galaxy$ & $M(M_{\odot})$ &$D_{ol}(Mpc)$ & $\Delta T_{2,1}$ ($\mu=0,q=0$)  & $\Delta T_{2,1}$($\mu=0,q=0.3$)&$\Delta T_{2,1}$($\mu=3,g=0.2,q=0.3$) \\
\tableline 
M87 & $6.5\times 10^9 $&
$16.68$ & $17378.9$ & $16186.6$ & $21000.2$\\
NGC 4472 & $2.54\times 10^9 $& 16.72&6791.12&6325.3&8206.24\\
NGC 4395 & $3.6\times 10^5 $& 4.3&0.962522&0.896499&1.16309\\
NGC1332 & $1.47\times 10^9 $& 22.66&3930.3&3660.71&4749.28\\
NGC 7457 & $8.95\times 10^6 $& 12.53&23.9294&28.2880&28.9157\\
NGC 1399 & $8.81\times 10^8 $& 20.85&2355.5&2193.93&2846.34\\
NGC 1374 & $5.90\times 10^8 $& 19.57&1577.47&1469.26&1906.17\\
NGC 4649 & $4.72\times 10^9 $& 16.46&12619.7&11754.1&15249.4\\
NGC 3607 & $1.37\times 10^8 $& 22.65&366.293&341.168&442.62\\
NGC 4459 & $6.96\times 10^7 $& 16.01&186.088&173.323&224.864\\
NGC 4486A & $1.44\times 10^7 $& 18.36&38.5009&35.86&46.5236\\
NGC 1316 & $1.69\times 10^8 $& 20.95&451.85&420.857&546.006\\
NGC 4382  & $1.30\times 10^7 $& 17.88&34.7577&32.3736&42.0004\\
NGC 5077 & $8.55\times 10^8 $&
38.7&2285.9&2129.9&2762.34\\
NGC 7768  & $1.34\times 10^9 $& 116.0&3582.72&3336.97&4329.28\\
NGC 4697  & $2.02\times 10^8 $& 
12.54&546&503.036&652.622\\
NGC 5128  & $5.69\times 10^7 $& 3.62&152.132&141.697&183.833\\
NGC 5576  & $2.73\times 10^8 $& 25.68&729.912&679.845&882.009\\
NGC 3608  & $4.65\times 10^8 $& 22.75&1243.26&1157.98&1502.32\\
M32  & $2.45\times 10^6 $&
0.806&6.5505&6.10118&7.91547\\
Cygnus A  & $2.66\times 10^9 $& 242.7&7111.97&6624.13&8593.93\\
\hline
\hline
\end{tabular}
\end{table*}

\section{Comparison with observation:}\label{sec:comparison}

Commonly understood as a specialized form of modified Bardeen black holes, standard astrophysical entities such as ordinary regular Bardeen and Schwarzschild black holes have been extensively studied. Prior investigations have thoroughly explored their shadow casting and gravitational lensing. In this study, we expand on previous works conducted by researchers such as Bozza, Virbhadra, Ellis for the Schwarzschild black hole \cite{Bozza:2002zj, Virbhadra:1999nm, Bozza:2003cp}, He et al. \cite{He:2021htq}, and Stuchik and Schee \cite{Schee:2015nua, Stuchlik:2019uvf} focusing on regular Bardeen black holes, along with Islam et al.'s work \cite{islam2022strong} examining Bardeen black holes within 4D Einstein's gravity.

Islam et al. specifically investigated strong gravitational lensing by the Bardeen black hole in 4D Einstein's Gauss-Bonnet gravity, determining black hole parameters using data from supermassive black holes. Their research also highlights that the regular Bardeen black hole solution is a unique outcome of the Bardeen black hole within 4D Einstein's Gauss-Bonnet gravity. Recently, He et al. \cite{He:2021htq} scrutinized the shadow and observed properties of Bardeen black holes within diverse accretion models.

This paper focuses on discussing the shadow and strong gravitational lensing effects of the modified Bardeen black hole, juxtaposing it with the Schwarzschild ($\mu=0$ and $q=0$) and ordinary regular Bardeen ($\mu=0$) black holes.

In the shadow analysis, the numerical determination of the black hole's shadow radius is presented in Table.\ref{table:1}. It's evident that the radius of the modified Bardeen black hole's shadow exceeds that of both regular Bardeen and Schwarzschild black holes. The angular diameter of the modified Bardeen black hole's shadow, as a function of parameters ($\mu/8M^2$ and $q/2M$), is visualized in Fig.\ref{fig:2}. Interestingly, the angular diameter of the modified Bardeen black hole surpasses that of the regular Bardeen black hole, especially in relation to supermassive black holes like $M87^{*}$ and $Sgr A^{*}$ (refer to Fig.\ref{fig:2}). Moreover, within $1\sigma$ uncertainty, specific angular diameter values, such as $\theta_d=39.4615 \mu as$ for the modified Bardeen black hole, $\theta_d=38.80 \mu as$ for the Bardeen black hole, and $\theta_d=39.9265 \mu as$ for the Schwarzschild black hole, align with measured angular diameters ($\theta_d=42\pm3 \mu as$ for $M87^{*}$ and $\theta_d=51.8\pm 2.3 \mu as$ for $SgrA^{*}$).

Concerning strong gravitational lensing, we employ the method proposed by Bozza \cite{Bozza:2002zj} to differentiate various static spherically symmetric black holes, analyzing astrophysical outcomes through supermassive black holes (see Table.\ref{table:3} \& Table.\ref{table:5}) concerning modified Bardeen, ordinary regular Bardeen, and Schwarzschild black holes. Additionally, we estimate observable quantities such as angular position $\theta_{\infty}$, $S$, and $r_{mag}$ for these black holes, employing the supermassive black hole $NGC 4649$ with a mass of $M=4.72\times 10^9$ and a distance of $D_{OL=0.008}$ Mpc for numerical comparisons.

Our estimations, considering supermassive black holes with similar mass and distance, consistently reveal that the angular position of the innermost image $\theta_{\infty}$ and angular separation $S$ consistently surpass those of ordinary regular Bardeen and Schwarzschild black holes. Additionally, the modified Bardeen black hole showcases larger relative magnifications. Numerically, differences in $\theta_{\infty}$, $S$, and $r_{mag}$ between the modified Bardeen ($\mu=3$,$g=0.2$, $q=0.3$) and ordinary regular Bardeen ($\mu=0$, $q=0.3$) are approximately $5.9 \mu as$, $0.3 \mu as$, and $3.1$ magnitude, respectively. Likewise, differences between the modified Bardeen ($\mu=3$,$g=0.2$, $q=0.3$) and standard Schwarzschild black hole ($\mu=0$, $q=0$) are roughly $4.9 \mu as$, $0.18 \mu as$, and $2.2$ magnitude, respectively. Noteworthy ranges include $\theta_{\infty}\in (19.59,19.82)$, $S\in (0.00297,0.003138)$, and $r_{mag}\in (3.47,3.5)$ when $\mu=03$, $g=0.2$, and $0\leq |q|\leq 3$. These findings suggest that the outermost image for the modified Bardeen black hole is considerably closer to the innermost images, potentially allowing for separation from other black hole images.

In essence, the ability to detect the outermost relativistic image could facilitate distinguishing the modified Bardeen black hole from other standard astrophysical black holes like Schwarzschild and ordinary regular Bardeen black holes using technology. However, observationally, this proves challenging, given the angular separation of the relativistic images not exceeding $\sim 0.3 \mu as$. Moreover, the Einstein's ring $\theta^{E}_{\infty}$ for the modified Bardeen black hole is observed to be larger than that of other standard astrophysical black holes like Schwarzschild and ordinary regular Bardeen black holes (refer to Fig.\ref{fig:11}).

Additionally, analysis (refer to Table.\ref{table:4}) reveals that time delays between two relativistic images for a modified Bardeen black hole ($\sim 15249.4$ minutes) significantly exceed those of other astrophysical black holes such as the Schwarzschild black hole ($\sim 12619.7$ minutes) and ordinary regular Bardeen black holes ($\sim 11754.1$ minutes) in the context of the supermassive black hole $NGC 4649$. This discrepancy suggests that distinguishing the first and second relativistic images from observation may offer a better chance of differentiating the modified Bardeen black hole from other astrophysical black holes like Schwarzschild or ordinary regular Bardeen black holes. Consequently, a modified Bardeen black hole could be quantitatively distinguished from a Schwarzschild or ordinary regular Bardeen black hole.

The identification and confirmation of a modified Bardeen black hole, if achieved, would bear several substantial implications and consequences for our understanding of black holes and general relativity. Some potential implications include:
\begin{itemize}
    \item Providing empirical evidence for alternative theories of gravity beyond the predictions of general relativity.
\end{itemize}
\begin{itemize}
    \item Displaying distinct astrophysical signatures and observational features compared to standard black holes, deepening our comprehension of their underlying physics.
\end{itemize}
\begin{itemize}
\item  Offering insights into the nature of dark matter, potentially constraining modified gravity theories that aim to explain dark matter effects.
\end{itemize}
\begin{itemize}
    \item Challenging the no-hair theorem in general relativity if the modified Bardeen black hole exhibits additional parameters or properties beyond mass, charge, and angular momentum.
\end{itemize}
\begin{itemize}
    \item Pushing the boundaries of our understanding of fundamental physics, inspiring new theoretical investigations and potentially leading to the development of more comprehensive theories explaining black hole behavior in modified gravity scenarios.
\end{itemize}

\begin{table*}
 \caption{Estimation of observables by taking the supermassive black hole $NGC 4649$  having mass $M=4.72\times 10^9 M_{\odot}$ and distance $D_{OL}=16.46 Mpc$ in the context of  Schwarzschild, ordinary regular Bardeen,and modified Bardeen black hole spacetimes.
\label{table:5}}
\begin{tabular}{p{1.5cm}|p{2.1cm}| p{6.5cm}|p{6.5cm}| }
\hline
\hline
\multicolumn{4}{c}{Comparison with observables}\\
\hline
 & Schwarzschild BH & Regular Bardeen BH ($\mu=0$) & Modified Bardeen BH ($\mu=3, g=0.2$) \\
   & $(\mu=0,q=0)$ & $|q|=0.0$ $|q|=0.1$ $|q|=0.2$ $|q|=0.4$ & $|q|=0.0$ $|q|=0.1$ $|q|=0.2$ $|q|=0.4$ \\
 \tableline 
$\theta_{\infty}(\mu as)$ & 14.6901 &$14.6656$~~~~$14.591$~~~$14.27$~~~$13.6825$ & $19.8192$~~$19.7998$~~$19.7227$~$19.5959$ \\
 \tableline 
S$(\mu as)$ & 0.0184 &$0.018$~~$0.0196$~~$0.024$~~$0.037$ & $0.00313$~~$0.00312$~~$0.00307$~~$0.00297$ \\
 \tableline 
$r_{mag}$& 6.82188 &$6.80282$~~$6.74425$~~$6.48595$~~$5.92957$ & $8.95283$~~$8.95941$~~$8.98787$~~$9.04274$  \\
\tableline 
$u_{ph}/R_{sh}$ & 2.59808 &$2.59373$~~$2.58054$~~$2.52504$~~$2.41987$ & $3.5052$~~$3.50177$~~$3.48813$~~$3.4657$  \\
\tableline 
$\bar{a}$ & 1.00 &$1.0028$~~$1.01151$~~$1.05179$~~$1.15041$ & $0.761981$~~$0.761421$~~$0.75901$~~$0.754404$  \\
\tableline 
$\bar{b}$& -0.40023 &$-0.401608$~~$-0.406055$~~$-0.429858$~~$-0.511002$ & $-0.384548$~~$-0.381893$~~$-0.37101$~~$-0.35205$  \\
\hline
\hline
\end{tabular}
\end{table*}

\section{Results and Conclusions}\label{sec:results}

In this study, we have comprehensively explored the observational characteristics of the modified Bardeen black hole through shadow and strong gravitational lensing observations, contrasting these features with those of other astrophysical black holes like the Schwarzschild black hole and the ordinary regular Bardeen black hole. Our analysis delved into understanding how the parameters associated with the modified Bardeen black hole influence its observable shadow and strong lensing effects.

To commence, we derived the null geodesics pertinent to the modified Bardeen black hole using the Hamiltonian-Jacobi action and conducted an extensive review of the photon orbit encompassing this specific black hole. Employing numerical techniques, we estimated both the photon sphere radius and shadow radius. Notably, we observed a distinct trend: for fixed values of the parameters $\mu$ and $g$, these radii exhibited a decrease with an increase in the magnitude of the charge parameter $q$. Conversely, for fixed values of $q$ and $g$, these radii displayed an increase with a rise in the magnitude of $\mu$. Furthermore, our investigation revealed that the shadow radius associated with the modified Bardeen black hole is notably larger than that of both the Schwarzschild black hole and the ordinary regular Bardeen black hole.

Our study also involved determining the angular diameter of the black hole shadow concerning the parameters $\mu$ and $q$ for a fixed parameter $g$ (specifically, $g=0.2$). We specifically considered this parameter setup in relation to supermassive black holes such as $M87^*$ and $SgrA^*$. Our analysis vividly demonstrates that the angular diameter of the black hole shadow attributed to the modified Bardeen black hole surpasses that of both the Schwarzschild and regular Bardeen black holes. Moreover, employing the EHT (Event Horizon Telescope) collaboration's data on the angular shadow diameter of $M87^*$ and $SgrA^*$, we were able to confine the parameter ranges of $\mu$ and $q$ for the modified Bardeen black hole. Our observed constrained ranges are as follows: $-0.89\leq \mu/8M^2 \leq 0.4$ and $0\leq |q|\leq 0.185$ for $M87^*$; and $-1.38\leq \mu/8M^2 \leq 0.1$ and $0\leq |q|\leq 0.058$ for $SgrA^*$, maintaining the fixed value of $g/2M=0.2$.

These findings highlight the viability of modified Bardeen black holes as potential astrophysical black hole candidates, notably incorporating additional parameters $\mu$, $g$, and $q$ alongside the black hole's mass $M$, akin to the supermassive black holes $M87^*$ and $SgrA^*$. Moreover, the constraints derived from the EHT data significantly narrow down the parameter space ($\mu$, $q$) for the modified Bardeen black hole, adding credence to its astrophysical relevance.

Subsequently, we went through an extensive examination of the strong gravitational lensing phenomena caused by the modified Bardeen black hole, shedding light on its consequential astrophysical implications. Our focus centered on scrutinizing the influence of the parameters $\mu$, $g$, and $q$ associated with the modified Bardeen black hole on the strong deflection angle and other crucial strong lensing observables.

Our investigative approach involved a meticulous reexamination of the null geodesic equations, coupled with numerical estimations to ascertain the photon radius, ultimately leading to the derivation of the lensing coefficients $\bar{a}$, $\bar{b}$, and $u_{ph}/R_{sh}$. The insights garnered from our analyses revealed intriguing trends: when considering fixed values of $\mu$ and $g$, we observed that $\bar{a}$ exhibited an increase while $\bar{b}$ displayed a decrease with the escalating magnitude of the charge parameter $q$. Conversely, for fixed values of $q$ and $g$, $\bar{a}$ depicted a decrease while $\bar{b}$ showcased an increase with the elevation of the parameter $\mu$. Notably, our observations regarding the deflection angle $\alpha_D$ unveiled a nuanced pattern wherein it initially experiences a slight increase, achieves a maximal value, and subsequently diminishes with the rising magnitude of the charge parameter $q$, given fixed values of $\mu$ and $g$. However, the deflection angle consistently diminishes with increasing $\mu$, maintaining fixed values of $q$ and $g$. Notably, in comparison to other astrophysical black holes like the Schwarzschild black hole and the ordinary regular Bardeen black hole, the deflection angle associated with the modified Bardeen black hole manifests as relatively smaller.

Furthermore, we conducted numerical estimations to determine the strong lensing observables pertinent to the relativistic images, specifically focusing on the modified Bardeen black hole within the context of supermassive black holes such as $M87^*$, $SgrA^*$, and $NGC7457$. Our findings unveiled notable disparities: the angular position $\theta_\infty$ and magnification $r_{\text{mag}}$ associated with the relativistic images for the modified Bardeen black hole surpassed those observed for both the Schwarzschild black hole and the ordinary regular Bardeen black hole. However, it's worth noting that the angular separation $S$ between the relativistic images concerning the modified Bardeen black hole was relatively smaller compared to the Schwarzschild black hole and the ordinary regular Bardeen black hole. To offer more precise insights, we outlined specific numerical ranges for these observables, $\theta_\infty$ and $S$, accommodating various values of the parameters $\mu$, $g$, and $q$, considering the gravitational influence of the supermassive black holes $M87^*$, $SgrA^*$, and $NGC7457`$. In the cases, where $\mu=0$ and
$0\leq |q|\leq 0.4$,$\theta_{\infty}\in (17.06,19.97)\mu as$ for
$M87^{*}$,$\theta_{\infty}\in (22.5,26.3)\mu as$ for $SgrA^{*}$
and $\theta_{\infty}\in (0.031,0.037)\mu as$ for $NGC 7457$; when
$\mu=1$ ,$g=0.2$ and $0< |q|\leq 0.4$,$\theta_{\infty}\in
(21.5,22.8)\mu as$ for $M87^{*}$,$\theta_{\infty}\in (28.3,30)\mu
as$ for $SgrA^{*}$ and $\theta_{\infty}\in (0.03,0.042)\mu as$ for
$NGC 7457$;  when $\mu=1$ ,$g=1.2$ and $0< |q|\leq
0.4$,$\theta_{\infty}\in (19.2,21.44)\mu as$ for
$M87^{*}$,$\theta_{\infty}\in (25.3,28.3)\mu as$ for $SgrA^{*}$
and $\theta_{\infty}\in (0.35,0.4)\mu as$ for $NGC 7457$; when
$\mu=3$ ,$g=0.2$ and $0< |q|\leq 0.4$,$\theta_{\infty}\in
(26.3,26.94)\mu as$ for $M87^{*}$,$\theta_{\infty}\in
(34.8,35.53)mu as$ for $SgrA^{*}$ and $\theta_{\infty}\in
(0.048,0.05)\mu as$ for $NGC 7457$; and  when $\mu=1$ ,$g=0.2$ and
$0< |q|\leq 0.4$,$\theta_{\infty}\in (23.6,24.7)\mu as$ for
$M87^{*}$,$\theta_{\infty}\in (31.19,32.58)\mu as$ for $SgrA^{*}$
and $\theta_{\infty}\in (0.048,0.05)\mu as$ for $NGC 7457$.
Moreover, the angular separation $S\in (0.024,0.14)\mu as$ for
$M87^{*}$, $S\in (0.032,0.184)\mu as$ for $SgrA^{*}$ ,$S\in (4.5
\times 10^{-5},2.54 \times 10^{-4})\mu as$ for  NGC $7457 $ for
the case when $\mu=0$ and $0\leq|q|\leq 0.4$; $S\in
(0.014,0.019)\mu as$ for $M87^{*}$, $S\in (0.019,0.025)\mu as$ for
$SgrA^{*}$ ,$S\in (2.6 \times 10^{-5},3.5 \times 10^{-5})\mu as$
for  NGC $7457 $ for the case when $\mu=1$ ,$g=0.2$ and $0<|q|\leq
0.4$;
 $S\in (0.03,0.19)\mu
as$ for $M87^{*}$, $S\in (0.04,0.16)\mu as$ for $SgrA^{*}$
,$S\in (5.5 \times 10^{-5},2.18 \times 10^{-4})\mu as$ for  NGC $7457 $ for the case when
$\mu=1$,$g=1.2$ and $0<|q|\leq 0.4$;
$S\in (0.003,0.0043)\mu
as$ for $M87^{*}$, $S\in (0.005,0.0057)\mu as$ for $SgrA^{*}$
,$S\in (6.9 \times 10^{-6},7.8 \times 10^{-6})\mu as$ for  NGC $7457 $ for the case when
$\mu=3$,$g=0.2$ and $0<|q|\leq 0.4$;
$S\in (0.022,0.025)\mu
as$ for $M87^{*}$, $S\in (0.02,0.033)\mu as$ for $SgrA^{*}$
,$S\in (4 \times 10^{-5},6.02 \times 10^{-5})\mu as$ for  NGC $7457 $ for the case when
$\mu=3$,$g=1.2$ and $0<|q|\leq 0.4$. Considering the supermassive black holes $M87^{*}$ and $SgrA^{*}$, the outermost Einstein's ring  $\theta^E_n$ has been displayed in Fig.\ref{fig:11} for the cases of modified Bardeen ($\mu=3,g=0.2$)  as well as ordinary regular Bardeen ($\mu=0$)  black holes.

Our analysis has notably unveiled that the outermost Einstein rings, denoted as $\theta^E_n$, associated with the modified Bardeen black hole are significantly larger in comparison to those observed with the ordinary regular Bardeen black hole. This crucial distinction implies that the modified Bardeen black hole showcases a notably wider angular separation between multiple relativistic images. This increased separation not only enhances its detectability but also serves as a distinguishing factor setting it apart from the characteristics exhibited by the ordinary regular Bardeen black hole.

Moreover, in our exploration of various supermassive black holes, we conducted an investigation into the time delay between the first and second-order relativistic images pertaining to the modified Bardeen, ordinary Bardeen, and Schwarzschild black holes. Our findings revealed a substantial contrast, with the time delay for the modified Bardeen black hole being approximately $\sim 15249.4$ minutes. This duration notably surpasses the time delays observed for the Schwarzschild black hole ($\sim 12619.7$ minutes) and the ordinary regular Bardeen black holes ($\sim 11754.1$ minutes) within the context of the supermassive black hole $NGC 4649$. This substantial disparity suggests that the modified Bardeen black hole presents distinctive temporal signatures, offering a promising avenue for its potential detection and differentiation from other prevalent astrophysical black holes.

\section*{Acknowledgements}

N.U.M would like to thank  CSIR, Govt. of
India for providing Senior Research Fellowship (No. 08/003(0141))/2020-EMR-I).

\bibliographystyle{apsrev4-2}%
\bibliography{rm.bib} 

\end{document}